%% file: main.tex
\documentclass[manuscript]{acmart}

\input{macros}

\AtBeginDocument{%
  }

\setcopyright{acmlicensed}
\copyrightyear{2025}
\acmYear{2025}
\begin{document}

\title{A Longitudinal Study on Different Annotator Feedback Loops in Complex RAG Tasks}

\author{Sara Rosenthal}
\authornote{Both authors contributed equally to this research.}
\email{sjrosenthal@us.ibm.com}
\author{Maeda Hanafi}
\authornotemark[1]
\email{maeda.hanafi@ibm.com}
\affiliation{%
  \institution{IBM}
  \country{USA}
}

\author{Yannis Katsis}
\affiliation{%
  \institution{IBM}
    \country{USA}
  }

\author{Lucian Popa}
\affiliation{%
  \institution{IBM}
    \country{USA}
}

\author{Marina Danilevsky}
\affiliation{%
  \institution{IBM}
    \country{USA}
}

\renewcommand{\shortauthors}{Hanafi et al.}

\begin{abstract}  
\input{abstract}
\end{abstract}


\maketitle

\input{intro_new}

\input{relatedworks}

\input{preliminaries}

\input{tbl/survey_demographics}

\input{method}
\input{analysis}
\input{tbl/survey_internal_anno}

\input{discussion}

\input{conclusion}

%\begin{figure}[h]
%  \centering
%  \includegraphics[width=\linewidth]{sample-franklin}
%  \caption{1907 Franklin Model D roadster. Photograph by Harris \&
%    Ewing, Inc. [Public domain], via Wikimedia
%    Commons. (\url{https://goo.gl/VLCRBB}).}
%  \Description{A woman and a girl in white dresses sit in an open car.}
%\end{figure}

\begin{acks}
We would like to thank our internal and external annotators for their high-quality work in creating conversations using RAGAPHENE and participating in our surveys: Mohamed Nasr, Joekie Gurski, Tamara Henderson, Hee Dong Lee, Roxana Passaro, Chie Ugumori, Marina Variano, and Eva-Maria Wolfe and Defined.AI. We would like to thank Kshitij Fadnis for his work developing RAGAPHENE.
\end{acks}

\bibliographystyle{ACM-Reference-Format}
%\bibliography{sample-base}
\bibliography{references}

%\appendix

\end{document}

%% file: macros.tex
\usepackage{xspace}
\usepackage{multicol}
\usepackage{multirow}
\usepackage{makecell}
\usepackage{subcaption}
\usepackage{tcolorbox}
\usepackage{soul}

\newcommand{\quotes}[1]{``#1''}

\newcommand{\studyduration}{one-year period\xspace}

\newcommand{\ragspelledout}{retrieval-augmented generation\xspace}
\newcommand{\ragspelledoutcaps}{Retrieval-augmented generation\xspace}

\newcommand{\skilllevel}{level of practice\xspace}

\newcommand{\workbench}{\textsc{{RAGAPHENE}}\xspace}

\newtcbox{\yellowhighlight}[1][]{%enhanced,
nobeforeafter,
hbox,
tcbox raise base,
sharp corners, 
 colback=yellow,
 colframe=gray,
 size=fbox,
 #1}

\newtcbox{\inlinebox}[1][]{%enhanced,
box align=base,
nobeforeafter,
sharp corners, 
 colback=yellow,
 colframe=gray,
 size=small,
 left=0pt,
 right=0pt,
 boxsep=2pt,
 boxrule=0.5pt,
 text height=1ex
 #1}

\newcommand{\slightlyfamiliar}{\inlinebox{ } \inlinebox{ }}
\newcommand{\somewhatfamiliar}{\inlinebox{ } \inlinebox{ } \inlinebox{ }}
\newcommand{\veryfamiliar}{\inlinebox{ } \inlinebox{ } \inlinebox{ } \inlinebox{ }}
\newcommand{\extremelyfamiliar}{\inlinebox{ } \inlinebox{ } \inlinebox{ } \inlinebox{ } \inlinebox{ }}

\newcommand{\beginner}{\inlinebox{ }}
\newcommand{\advancedbeginner}{\inlinebox{} \inlinebox{}}
\newcommand{\competent}{\inlinebox{ } \inlinebox{ } \inlinebox{ }}
\newcommand{\proficient}{\inlinebox{ } \inlinebox{ } \inlinebox{ } \inlinebox{ }}
\newcommand{\expert}{\inlinebox{ } \inlinebox{ } \inlinebox{ } \inlinebox{ } \inlinebox{ }}

%% file: abstract.tex
Grounding conversations in existing passages, known as Retrieval-Augmented Generation (RAG), is an important aspect of Chat-Based Assistants powered by Large Language Models (LLMs) to ensure they are faithful and don't provide misinformation. Several benchmarks have been created to measure the performance of LLMs on this task. We present a longitudinal study comparing the feedback loop of an internal and external human annotator group for the complex annotation task of creating multi-turn RAG conversations for evaluating LLMs. We analyze the conversations produced by both groups and provide results of a survey comparing their experiences. Our study highlights the advantages of each annotator population and the impact of the different feedback loops; a closer loop creates higher quality conversations with a decrease in quantity and diversity. Further, we present guidance for how to best utilize two different population groups when performing annotation tasks, particularly when the task is complex.

%% file: intro_new.tex
\section{Introduction}

%- Intro to RAG
Chat-based platforms such as ChatGPT \cite{OpenAI_ChatGPT} and Claude \cite{claude3} have become increasingly popular tools for asking questions and having conversations~\cite{lin2023review}. Given a question by a user, the task of grounding the answers in the conversation generated by the large language model (LLM) has become increasingly important to avoid hallucination or misinformation~\cite{huang2025survey}.  To address this problem, a considerable amount of research has been conducted in \ragspelledout (RAG) methods~\cite{lewis2020retrieval,gao2023retrieval}, which combine (1) a retriever model for fetching relevant passages from a corpus, with (2) a generator model to create a response to the user question based on the retrieved passages.
%which is a two-step setup that uses (1) a retriever model that retrieves relevant passages from a corpus to respond to questions in a conversation and (2) a generator model that generates a response to questions in a conversation using the passages retrieved from step 1. 

%- Evaluating RAG is important + explain the intuition for the HCI folks: 
A major aspect of evaluating a RAG method is understanding how well it handles \textit{challenging} and \textit{complex} questions, which are important properties of a \textbf{high-quality} %\footnote{We quantify what a high-quality conversation is further in Section \ref{sec:metrics}.} 
conversation. For example, can the RAG system find the correct documents to answer the human user's question, or else determine that giving a grounded answer is impossible? Can the RAG system handle vague user questions (\quotes{What time does the bank open?}), and successfully recover by, for example, responding by asking the user for clarification (\quotes{Which branch are you planning to visit?}).
%Intuitively, a conversation is challenging for the underlying RAG model, if .... \todo{explain the intuition}. 
%- Evaluating RAG models is thus important: 
Thus, there is a breadth of work on multi-turn RAG benchmarks~\cite{katsis2025mtragmultiturnconversationalbenchmark, dziri-etal-2022-faithdial, feng2021multidoc2dial, kuo2024radbenchevaluatinglargelanguage, es-etal-2024-ragas}.
%, which is especially important in evaluating and understanding the capabilities of a RAG setup. 
A high-quality benchmark for conversational RAG serves to illuminate the capabilities and boundaries of the underlying RAG method - in contrast with simpler conversations and easily answered turns.
%A high-quality benchmark for conversational RAG helps one understand the extent of the learned boundaries of the underlying RAG method as opposed to simpler conversations with turns that are relatively easy to answer with a simple or non-RAG method. 

%- What does this entail for annotators + annotation task (need to spell it out for them HCI folks): 
%To test the capabilities of the underlying RAG method, one has to create such benchmarking data. 
Human annotation is an important aspect of building high-quality benchmarks. 
%In this work, we tasked human annotators to create challenging and complex conversations for evaluating RAG models. 
%- RAG data is hard to create
However, manually creating conversational RAG data is hard; past research has shown it to be a cognitively demanding task even for experienced annotators \cite{ragdatahard}. 
The challenge lies in working with underlying retrieval and generation models to create a faithful and naturally flowing conversation (rather than, e.g., a set of disjoint individual questions and answers).
%An annotator must create a conversation, converse with the underlying RAG models, and ensure that the conversation flows naturally (and does not appear disjoint from the other turns). 
The annotator must drive the conversation by creating a thoughtful sequence of questions; on they other hand they must also edit the underlying RAG model's default-generated responses to be accurate and complete.
Finally, the annotator must also ensure that each turn is accompanied by one or more passages that are used to ground the information in the response. These aspects are all necessary to ensure the quality and complexity of the final conversation~\cite{chakrabarty2025aiwritingsalvagedmitigating, humanllmcollab, yao2025improvingsummarizationhumanedits}.
%- Challenging task: 
%The task is challenging; it has not been found in previous literature, and it is also challenging for the task requesters (the authors and researchers of this paper). Important to emphasize that the task is challenging, even for the task requesters, as the task has never been found before in the literature, i.e., you cannot be an expert in this task beforehand. There are no expert annotators on the task of creating conversational RAG data, in the sense that someone has had extensive experience completing such a task. 

%- Available options for hiring annotators: 
Common approaches for collecting human-annotated data vary from hiring professional annotation services to hiring crowd workers via platforms such as Mechanical Turk\footnote{https://www.mturk.com/} and Scale.AI\footnote{https://scale.com/}, with various tradeoffs in cost, worker reliability, and data quality ~\cite{klie2024analyzingdatasetannotationquality,snow-etal-2008-cheap, wang2024case, agreementnotgold2023, asher-etal-2016-discourse}. 
%- Introduce the real-world constraint + prior work: 
Most previous literature in crowdsourcing assumes a \textbf{direct communication feedback loop} between task requesters and annotators. However, in some real-world cases, task requesters cannot directly communicate with the annotators and  must go through an intermediary - a common arrangement when hiring external annotation services. 
% Teaser of prior work
Prior work has shown that different annotator communication strategies can impact the final dataset~\cite{shepherding2012, surveytaskassignment2022}; for instance, different annotator hiring mechanisms support various feedback loops of communication (e.g., direct/indirect) and frequency (e.g., weekly/daily).

%- Introduce the external and internal annotators:
In this work, we tasked human annotators to create challenging and complex conversations for evaluating RAG models. 
In particular, we worked with two groups of annotators, which we refer to throughout the paper as \textbf{internal} and \textbf{external}. 
The internal annotators were hired as part of the same organization as the researchers (task requesters); they have a \textbf{direct communication feedback loop} with the task requesters and an hourly pay rate. The external annotators were hired as part of a dedicated external crowdsourcing annotation service, and have an \textbf{indirect communication feedback loop}, where all communication---including task instructions, task guideline materials, and feedback and per task financial incentive---between the task requesters and annotators passes through an intermediary representative from the annotation service.

%In our work: 
In this paper, we present the results from a longitudinal study over an approximately \studyduration (May 2024 to May 2025), based on observations of \textbf{internal} and \textbf{external} annotators.
We instruct both groups to create complex and high-quality multi-turn conversations for RAG. This task is complex and hard~\cite{ragdatahard}, and it allows 
us to showcase both the challenges and the differences when employing direct and indirect \textit{feedback loops} between the task requester and annotators.
%in order to focus on the the challenges that arise in direct and indirect \textit{feedback loops} between the task requester and annotators for a task that is considered complex and hard~\cite{ragdatahard}.
%, and focused on the differences in communication feedback loops. 
%What are the challenges that arise in direct and indirect \textit{feedback loops} between the task requester and annotators for a task that is considered complex and hard~\cite{ragdatahard}? 
In particular, we 
%consider 
show 
how the different feedback loop structures 
%can potentially 
impact the quality of the created conversation data, and address the question of which strategies should be employed to improve final data quality.
%How do these different feedback loop structures potentially impact the quality of the data? What strategies should one employ to improve the quality of the final data? 
%- Clarify the focus of our work: 
%In this paper, we present the results from a longitudinal study throughout an approximately \studyduration (starting from May 2024 to May 2025) where we observed two such groups of annotators.
%We focus on the challenges that arise when we task two groups of annotators with different communication feedback loops, and we report our observations in the final created conversational RAG data from the two groups of annotators. 

\begin{figure}
    \centering
    \includegraphics[width=0.5\linewidth]{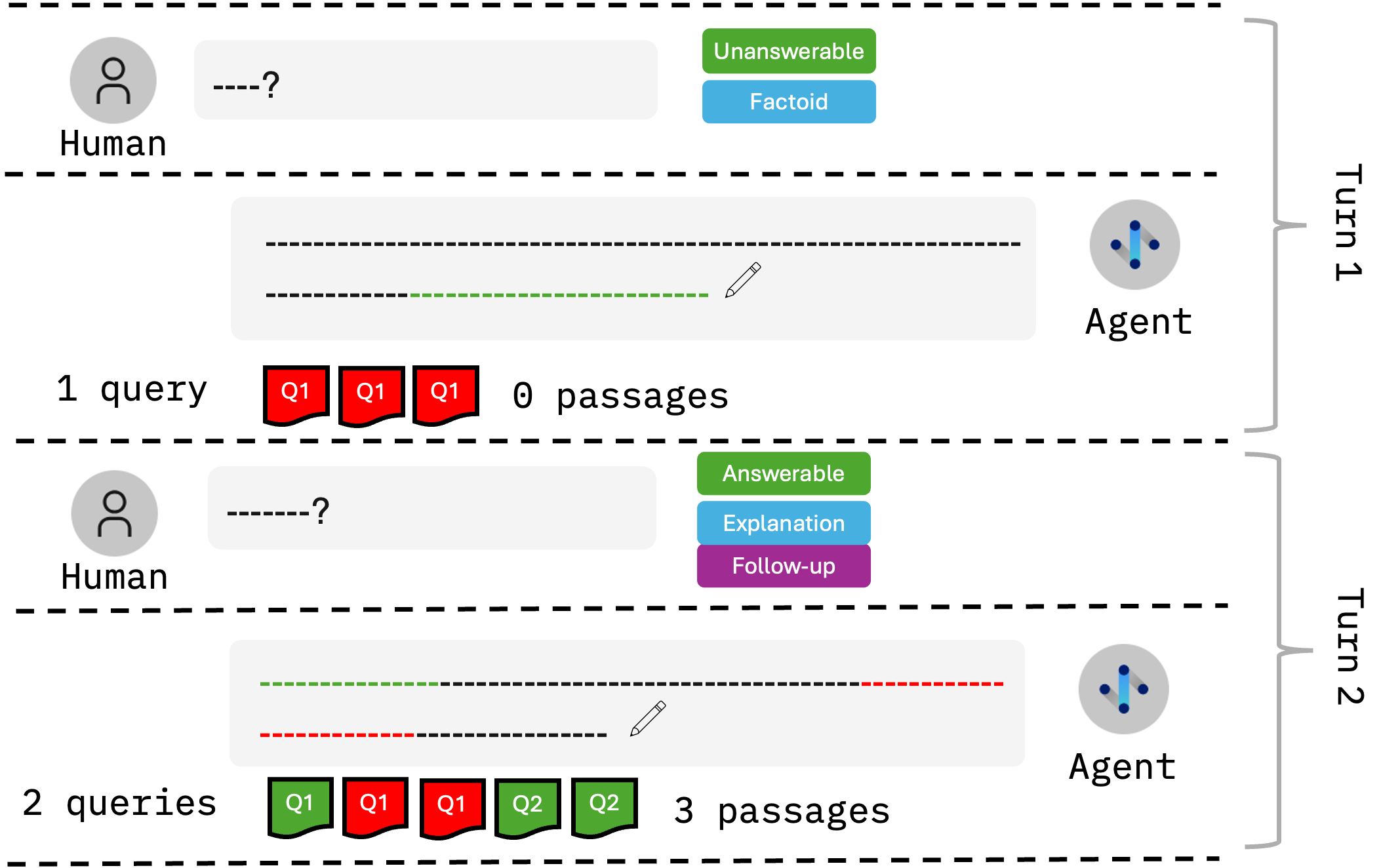}
    \caption{A sample of what the structure of a multi-turn RAG conversation looks like. This conversation has two turns (a question from the human followed by an answer from the AI conversational agent). The first question is unanswerable and has no relevant passages. The second question is answerable and has three relevant passages that were found from two different queries (Q1 and Q2). Both answers were edited from the original answer by adding (green) and/or removing (red) text.}
    \label{fig:sample_conv}
\end{figure}

% Extra deets on the task + data + tooling
The annotators interact with an AI conversational agent via our annotation tool called \workbench \cite{fadnis2025ragapheneragannotationplatform} that has been specifically designed to support building complex conversational datasets. An example of the resulting conversation structure is shown in Figure~\ref{fig:sample_conv}.
We report on the quality of the created conversations (quality is defined in Section \ref{sec:metrics}) for both annotator populations for three phases: pilot, creation, and review. We evaluate our findings using automated metrics, including the number of accepted/rejected conversations after manual review, and metrics that compare conversation properties. We also ask participating annotators to take a survey regarding their experiences, and compare the results between the two annotator groups in terms of their perceptions of both the tool and the task. Analyzing both data quality and annotator experience helps us  identify the most important aspects of the task and how to best utilize the two annotator groups. %to focus on in the quality assurance (QA) process. 

% \todo{Formulate the research questions here alongside contributions?}
Our contributions are as follows:
\begin{itemize}
    \item Insight on how different communication feedback loops impact annotator data quality from annotators, specifically on complex tasks (conversational data for RAG). We find a gap between the two groups of annotators, wherein annotators with a closer communication feedback loop created higher-quality data than annotators who did not have a close communication feedback loop. 
    \item Insight on how different annotator groups perceive the task and annotation tool via a comprehensive analysis of surveys regarding their experience. We find that there are differences in how the two groups of annotators perform the task and perceive the annotation tool; in particular, that annotators with a closer communication feedback loop spend more time on the task and have a more positive impression of the tool.
    \item Concrete strategies for developing a rigorous process of data creation to handle real-world constraints. Our suggestions can help mitigate downstream data quality issues while increasing quantity and variety, especially in cases where one group of annotators lacks the benefits of a closer communication feedback loop. 
\end{itemize}

% \todo{Sara: outline}

In the rest of this paper, we first describe related work regarding RAG for LLMs, data quality in human annotations, tooling methods, and human communication structure. In Section~\ref{sec:prelimanaries}, we describe and define the task of creating and reviewing conversations. We also define the metrics used to quantify the challenging and complexity level of the conversations created to ensure high quality. In Section~\ref{sec:method}, we describe our methods of communication and the phases of data collection. Finally, we end with a discussion of our insights on utilizing internal and external annotators in Section~\ref{sec:discussion}, and limitations and conclusions in Sections ~\ref{sec:limits} and ~\ref{sec:conclusion}.

%% file: relatedworks.tex
\section{Related Works}

There are four primary areas related to our work: \ragspelledout (RAG), data annotation quality, tooling methods for creating conversations and communication structures. Our work is unique in that we explore how real-world constraints on communication structures impact the quality of data annotation, specifically in the realm of complex tasks (in our case, creating multi-turn conversational RAG data).  

%Previous work in data creation and annotation spans includes focuses on tooling\cite{watsonxorders, dialoguedatacollection2015, ogawa-etal-2020-gamification, asher-etal-2016-discourse}, improving annotation guidelines\cite{}, and improving annotator and data quality\cite{klie2024analyzingdatasetannotationquality, snow-etal-2008-cheap, gpt4ascrowworker}.

\subsection{Retrieval-Augmented Generation (RAG) for LLMs}
\ragspelledoutcaps or \textbf{RAG}~\cite{rag2020, Soudani_2024} is a popular two-stage method for augmenting large language models with passage retrieval to generate grounded responses~\cite{llmshallucinations2024}. In a RAG setup, AI agents generate a response to a human query in a two-stage manner: (1) calling a \textbf{retrieval model} that returns documents relevant to the human query and (2) calling a \textbf{generative model} that generates a response given both the human query and the relevant documents. RAG is an especially popular method for enterprises to augment chatbots with up-to-date or domain-specific knowledge regarding a pre-existing corpus of documents~\cite{ragdatahard}. 

Existing work in RAG primarily focuses on the system components in a RAG method~\cite{es-etal-2024-ragas, lewis2020retrieval,gao2023retrieval}, and benchmarks such as multi-turn RAG datasets including RAD-Bench~\cite{kuo2024radbenchevaluatinglargelanguage}, RAGBench~\cite{friel2025ragbenchexplainablebenchmarkretrievalaugmented}, RGB~\cite{chen2023benchmarkinglargelanguagemodels}, and MTRAG~\cite{katsis2025mtragmultiturnconversationalbenchmark}. 
Benchmark datasets for RAG need to be high-quality and complex for the language model it is testing. Synthetically generating and augmenting conversational data is a popular method, however, they are not of the same quality as human-driven conversations~\cite{tan-etal-2024-large, seo2024retrievalaugmenteddataaugmentationlowresource, soudani2024surveyrecentadvancesconversational, papangelis-etal-2021-generative,mao-etal-2024-ragstudio}. Converser~\cite{huang-etal-2023-converser} generates conversations at a per-turn level, where the synthesized questions do not necessarily depend on each other in the same way naturally human-generated questions unfold in conversations.  Our findings in Section~\ref{sec:synthetic} on the quality of synthetic data are similar.  On the other hand, creating multi-turn conversational datasets is also challenging for human annotators~\cite{ragdatahard}. Previous existing papers that use humans to create multi-turn RAG conversations explore hired experts \cite{anantha2021opendomainquestionansweringgoes,adlakha2022topiocqaopendomainconversationalquestion}, crowdsourcing \cite{choi-etal-2018-quac, reddy-etal-2019-coqa, dinan2018wizard}, and hybrid \cite{katsis2025mtragmultiturnconversationalbenchmark} approaches but none have analyzed the impact of the different feedback loops on the quality of the data. We describe this process in detail in Section \ref{sec:task-create-conv}. Interviews with professional annotators show that the multi-layer aspect of creating such data results in a high cognitive load for users~\cite{ragdatahard, sweller1994cognitive}.

\subsection{Data Quality in Human Annotation} 

Improving the data quality from annotators is a well-studied area, including annotation quality management practices~\cite{klie2024analyzingdatasetannotationquality}, techniques involving Human-LLM collaboration~\cite{humanllmcollab, gpt4ascrowworker}, tradeoffs between cost and quality~\cite{snow-etal-2008-cheap}, etc. Most of these works are limited to \textit{micro-tasks}~\cite{snow-etal-2008-cheap, conv-entity-linking2021, crowdsourcemicrotaskconv}, which an annotator can complete quickly, requiring non-complex annotator input such as labeling an image or a sentence. 

Past research has pointed out the varying quality of data from different kinds of annotators \cite{expertfac2016,snow-etal-2008-cheap}. Crowd annotators, who are typically hired via crowdsourcing platforms like Mechanical Turk, can be tasked at cheaper rates and hired at a larger scale to create large labeled datasets very quickly. However, the data may contain noisy labels and not be up to par in terms of the quality needed for any downstream task \cite{snow-etal-2008-cheap}. 

Moreover, existing strategies for gathering high-quality annotation data from diverse crowd workers, such as comparing the gold standard to annotators' outputs~\cite{selectingannotators2015} or breaking up the task into MapReduce-like tasks~\cite{CrowdForge}, are not applicable for the complex task of creating conversational data for RAG. Creating a conversation is a mix of a creative task and a generative task, with the final output judged by metrics far different than simple accuracy or precision/recall~\cite{katsis2025mtragmultiturnconversationalbenchmark}.  The task of creating conversational data for RAG is essentially a \textit{non-decomposable macrotask}~\cite{nondecomposabletasks2018, surveytaskassignment2022}, where the task simply cannot be broken up and given to crowd workers in parallel to complete without degradation of the final conversation quality~\cite{ragdatahard}.

There have been past works ~\cite{wang2024case, agreementnotgold2023, asher-etal-2016-discourse, ensemble2014, shepherding2012} that tasked different groups of annotators of varying qualities.  In Wang et al.'s work~\cite{wang2024case, agreementnotgold2023}, an existing dataset was evaluated by two groups of annotators: an expert group and a more diverse \quotes{general} group, referred to as the \textit{crowd group}. While they found that expert annotators and the general annotators disagree on what parts of the data are considered safe in the context of AI safety, it is not clear what the impact of differing communication feedback structures between the two such annotator groups had on the final data. They observed differences in labeling between the two different groups of annotators and argued that it is not linked to a definition of accuracy (or an error) but rather a difference in domain expertise, where the expert annotators bring in-depth knowledge regarding institution specific policies and crowd annotators bring in diverse perspectives from their sociocultural background. We build on top of this notion to help us utilize the different strengths we identified in our two groups of annotators with different communication feedback structures to aid us in the complex and cognitively demanding task of creating conversational data for RAG. 

The usage of a multi-stage data creation process with human annotators has been used across many different applications. In Asher et al.'s work \cite{asher-etal-2016-discourse}, conversational data was collected during multi-player game sessions, and then, as a separate process, novice and expert annotators annotated the conversational data. However, this work did not study differences between the novice and expert annotators and their impact on the final data. In Lu et al.'s work ~\cite{crowdinnovation2016}, a two-stage process involving crowd annotators was used to complete design tasks. They showed that even if the crowd annotators do not possess a high level of expertise, they can search and utilize expert-generated resources to accomplish design tasks. However, this work is also limited to a certain kind of task and cannot easily be generalized to more complex tasks like creating conversational data for RAG. We take an extra step further with our exploration of a similar, multi-stage process involving our two groups of internal and external annotators: we explore in depth how our two groups of annotators, each with differing communication feedback structures impact the quality of the final data and how we overcame barriers introduced in different communication feedback structures.  

Ensemble~\cite{ensemble2014} proposes a framework for dividing certain tasks, such as providing goals, merging tasks, generating ideas, doing detailed edits, etc., that are assigned to either leaders or crowd collaborators. They studied a hierarchical type of communication structure between leaders and crowd collaborators for a creative type of task, specifically storytelling. However, their insights and their proposed framework for dividing up work cannot easily be extended to the task of creating conversational RAG data, as this task cannot be divided up into smaller pieces without compromising the quality of the whole conversation (such as disrupting the natural flow of the conversation).  We also acknowledge similar lines of work to Ensemble, that belong in the realm of \textit{online collaboration}, including similar tools for online collaborative document editing~\cite{collaborativewriting2012}, computational notebooks~\cite{computationnotebooks2019}, etc. Our tool \workbench~\cite{fadnis2025ragapheneragannotationplatform}, is, by nature, not meant to support online collaboration, but rather the online aspect is in the human communication with the agent as part of the process of creating a conversation. The different annotator groups collaborate on the same conversation object in an offline fashion, in a sequential manner, given the complexity involved in the task.

One technique we borrow from the literature is utilizing self-assessment and expert feedback \cite{shepherding2012, bostock2010crowdsourcedgraphical, annotationqualityassurance}, which has been proven to help annotators improve their work. We show how critical this strategy was in improving the quality of the created conversational RAG data.

%Previous work comparing crowdsourcing to LLM labeling: but our task is much more complex than the typical micro tasks that were studied before.
    
\subsection{Tooling Methods for Creating Conversational Data} 

There is a multitude of previous work involving tooling for creating and annotating conversational data, \cite{tilda2018, instantannotation2014, winder2021, dialoguedatacollection2015, ogawa-etal-2020-gamification}.
Some works have explored gamified ways of creating conversational data and annotating it, such as Manuvinakurike et al.'s work~\cite{dialoguedatacollection2015}, where they use a gamified crowdsourcing approach to gather conversational data, involving utterances between the players. However, their gaming platform cannot be easily extended to more complex forms of data structures (such as in our conversational RAG dataset) since the gameplay involves players identifying a target image amongst other images displayed on the screen.  

Ogawa et al. uses Minecraft\footnote{\url{https://www.minecraft.net/en-us}}, an online sandbox building game, as a way to gamify the process of creating such complex data structures~\cite{ogawa-etal-2020-gamification}. However, the conversational data created from this tool is a simpler structure than one involving RAG, as it consists of utterances between the players, and each utterance may be assigned classification labels.  A social aspect is integrated into the system via a ranking system of the players to incentivize players to annotate the conversational data. We also borrow these ideas and explore the usage of annotators commenting on each other's work to learn and improve the final quality of the created and annotated data (see Section \ref{sec:task-review-conv}). 

Moreover, past work has not yet explored the creation and annotation of conversational data for RAG, which is more complex given that a set of relevant passages is attached at each of the AI agent's turns and the answers must be faithful to those passages. The added complexity at the agent's turn has been shown to bring annotators to a state of cognitive overload~\cite{sweller1994cognitive} while creating and annotating conversation data for RAG ~\cite{ragdatahard}.

\subsection{Human Communication Structure}
% Existing work related to communication and conversation has been done across many contexts, including sociolinguistic study on instant messaging~\cite{sociolinguisitc2002}, usability studies of how best to present conversational data ~\cite{usabilityim2010, watsonxorders}, and tooling for annotating conversation, \cite{tilda2018, instantannotation2014, winder2021}. Conversation in such works is seen as the communication medium for collaboration, with annotations playing a role in facilitating and improving the communication. However, in our case, an annotated conversation is the output artifact, which requires our annotators to play the role of both creating a conversation and annotating it.

Prior work explores the impact of different communication structures on a process, such as Cataldo and Ehrlich's study on communication between coworkers in product development~\cite{longitudinal-productdev} or Erete and Burrell's study on communication between citizens and leaders in local government participation~\cite{longitudinal-gov}. In both of these works, a longitudinal study was employed to capture how different communication structures between people impact certain outputs in the process. In Erete and Burrell's study of communication and the role of technology across three different local communities, with different communication of email, discussion boards or instant messaging, they observed that technology use helps citizens amplify their concerns and bring accountability to local leaders, but it does not necessarily increase their political power. Cataldo and Ehrlich's work studied hierarchical structures (manager $\rightarrow$ report) versus smaller structures (peer $\leftrightarrow$ peer), where they found that hierarchical communication structures have a bigger impact on product delivery than smaller communication structures. While these insights from the different longitudinal studies are useful for their particular niches, it is not clear how these insights extend to the impact of data quality when the annotator groups have different communication feedback structures (direct or through an intermediary).  

There is extensive research on the concept of feedback in the HCI literature, including teacher-to-student feedback in classroom settings~\cite{bayerlein2014students, cutumisu2018impact, autofeedback2024}, methods for improving and reusing feedback from novices~\cite{reusingfeedback2018, aggregaringfeedback2015, almostexpert2016}, peer-review feedback~\cite{juxtapeer2018, PuzzleMe2021}, and feedback between peers and experts~\cite{cho2006commenting, fruitfulfeedback2017}. While many of these works use the term feedback to mean improving one's understanding of a particular task, none of these works explore what it means when that feedback is disjoint, as in our case with our external annotators, where all communication is done through an organization's intermediary. In our work, we show how differences in feedback loop structures have a significant impact on the final output data.

%% file: preliminaries.tex
\section{Tasks}
\label{sec:prelimanaries}

\begin{figure}
    \centering
    \begin{subfigure}[t]{0.60\textwidth}
        \centering
        \includegraphics[width=0.95\linewidth]{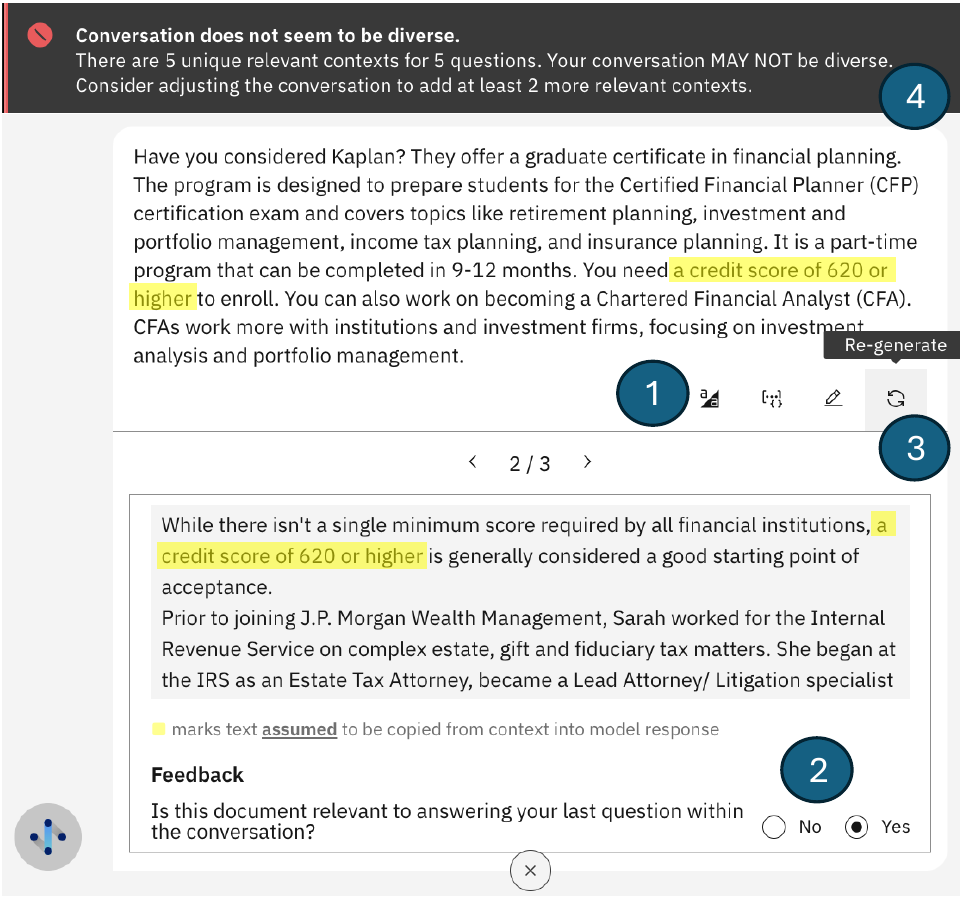}
        \caption{A view of the agent's response. The overlap button (1) highlights texts that overlap between the passage and the agent's response in yellow. (2) The annotator must mark passages as relevant/irrelevant to the turn. (3) The annotator can click on the regenerate button after changing the passages (adding or deleting passages from requerying). (4) Hints appear at the top of the conversation if it detects quality issues such as passage diversity, passage relevancy labels, missing enrichments, low number of edits, etc.  }
        \label{fig:editing-retriever-outputs}
    \end{subfigure}
    \hspace{1em}
    \begin{subfigure}[t]{0.34\textwidth}
        \centering
        \includegraphics[width=0.95\linewidth]{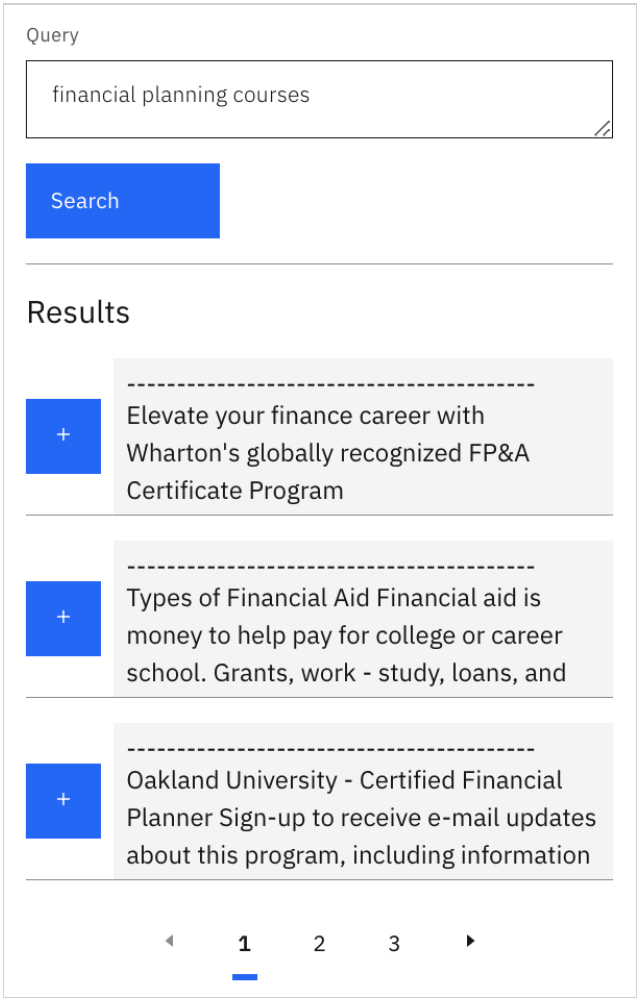}
        \caption{Requery tool for finding passages to attach to the turn. }
        \label{fig:requery}
    \end{subfigure}
    \\
    \begin{subfigure}[t]{0.50\textwidth}
        \centering
        \includegraphics[width=0.95\linewidth]{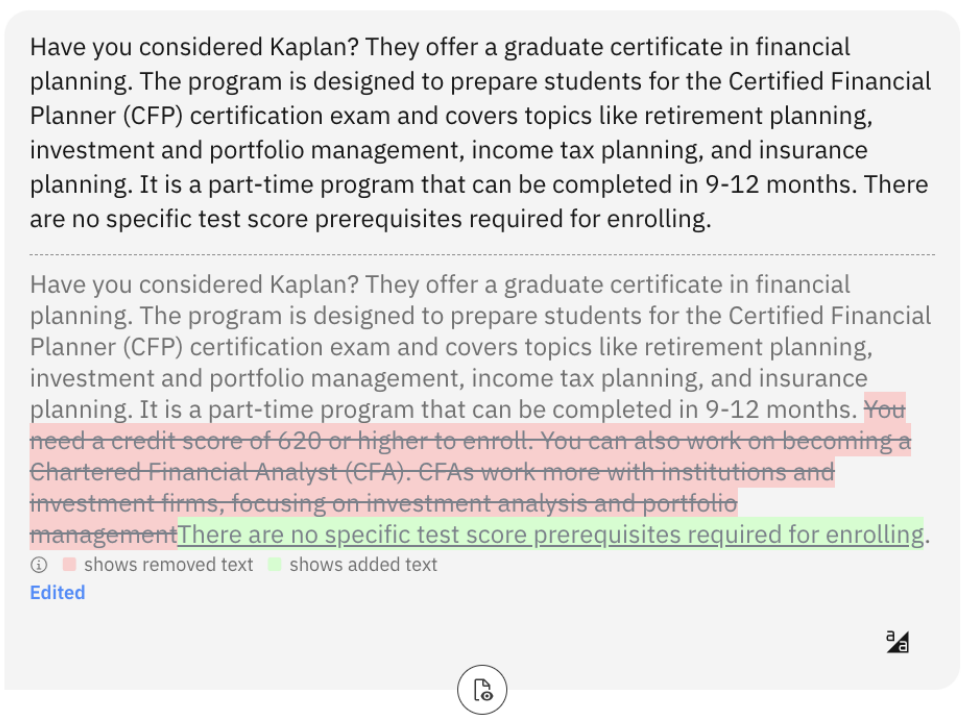}
        \caption{The annotator edits the agent's response (top half is the edited response). When the annotator edits the agent's response, a visualization of the differences between the original agent's response and the edits is shown (bottom half contains differences).}
        \label{fig:editing-generator-outputs}
    \end{subfigure}
    \begin{subfigure}[t]{0.44\textwidth}
        \centering
        \includegraphics[width=0.95\linewidth]{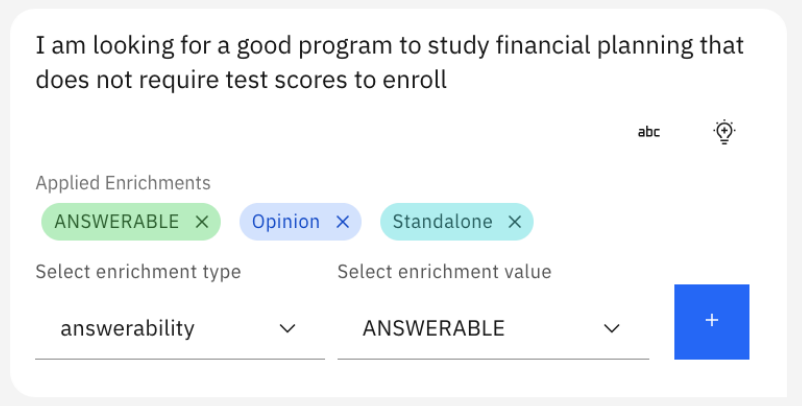}
        \caption{Enrichments on the user's text.}
        \label{fig:enrichments}
    \end{subfigure}
    %\begin{subfigure}[t]{0.49\textwidth}
    %    \centering
    %    \includegraphics[width=0.95\linewidth]{img/hints.pdf}
    %    \caption{Hints }
    %    \label{fig:hints}
    %\end{subfigure}
    \caption{View of creating a conversation in \workbench, when an annotator creates an initial message, \quotes{I am looking for a good program to study financial planning that does not require test scores to enroll}.}
    \label{fig:tool-features}
\end{figure}

\begin{figure}
    \centering
    \includegraphics[width=0.50\linewidth]{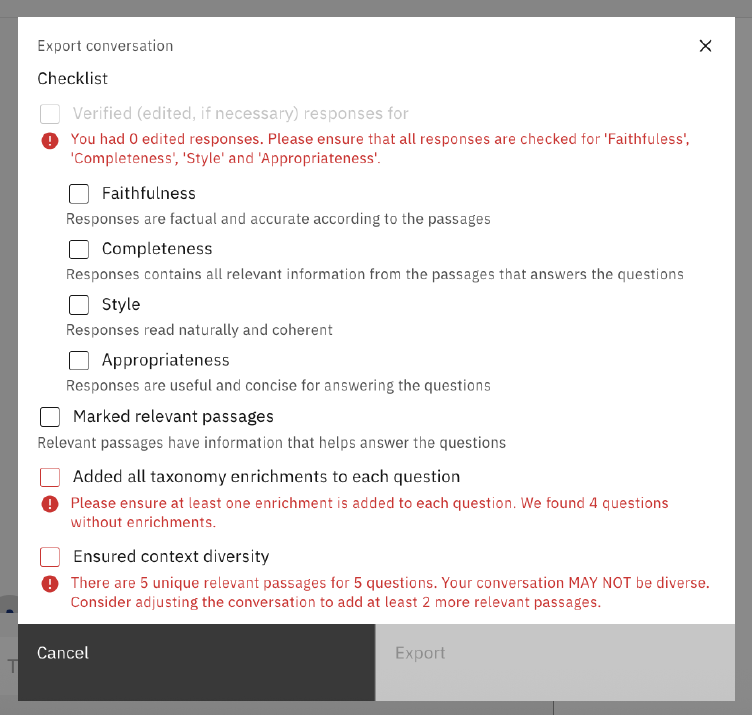}
    \caption{Checklist that an annotator has to go through before exporting the created conversation. }
    \label{fig:export-checklist}
\end{figure}

% \todo{From yannis: It would help to introduce RAG in a slower, more complete fashion. For instance, it would help to talk about the underlying document corpus (which currently is implied) and also explain the combination of retriever and generator that is employed in such systems (which currently only appears as part of the conversation generation task in Section 3.1). Maeda: Already rewrote intro to really walk the HCI audience through intuition of RAG and RAG evaluations}

\textbf{Multi-Turn Retrieval-Augmented Generation (RAG)} is the task of grounding an AI agent's response to each human input in the conversation using relevant passages. This is achieved by accessing relevant passages via retrieval of the provided corpus. The corpus can be the entire web, or a smaller domain-specific corpus that focuses on one topic (e.g. finance or tech). 
%In a \textbf{multi-turn RAG conversation}, a human will will ask questions to an AI agent. The agent answers the human by retrieving relevant \textbf{passages} or contexts to generate and ground the AI agent's response.
%consists of several alternating questions and answers between an AI agent and a human, where each AI agent can  have \textbf{contexts} or documents attached to the turn, by which the RAG component utilizes to generate and ground the AI agent's response. 
Each question-passages-answer pair asked by the human, retrieved from the corpus, and answered by the agent is considered a \textbf{RAG turn} in a conversation.
%comes from either a human user or an AI agent. In our case, a conversation for RAG consists of alternating turns between a human and an AI agent. 
The general structure of a RAG conversation can be found in Figure \ref{fig:sample_conv}.

% \todo{From yannis: need a more progressive introduction of RAGAPHENE and explain why we selected this particular platform (given that in a double-blind submission, this looks like somebody else’s work).}

The primary steps involved in creating a RAG conversation given to our annotators are: (1) creating a conversation and (2) reviewing a conversation. \workbench ~\cite{fadnis2025ragapheneragannotationplatform} is a chat-based annotation tool that enables users to simulate real-world multi-turn conversations. In addition to the ability to create and review conversations it has several key features that made it ideal for our use case: real-time retrieval and generation and the ability to edit responses. We describe how these tasks are accomplished using \workbench in the following sections. % \workbench was built for the task of creating Multi-Turn conversations. 
Screenshots describing how we utilized the annotation tool are shown in Figure~\ref{fig:tool-features}.

\subsection{Task: Creating conversational datasets for RAG}\label{sec:task-create-conv}

%The task of creating Multi-Turn RAG conversations is a complex and time-consuming task \cite{ragdatahard}. For each turn, the annotator must 1) Write a question, 2) Receive an agent response and relevant passages, 3) Edit the retrieved passages, 4) Edit the agent answer, and 5) Add labels and enrichments. This can take 30-60 minutes per conversation for an experienced annotator. \maeda{to add a walkthrough of creating a conversation, add an image showing an example of a conversation}

The task of creating multi-turn RAG conversations is a complex and time-consuming task \cite{ragdatahard}. The process for creating a single high-quality complex conversation occurs per turn. We describe how an annotator, Alice, creates a single turn for a conversation: 
\begin{enumerate}
    \item \textbf{Create a question}: Alice writes a question as shown in Turn 1 of Figure \ref{fig:sample_conv}. An example of a question is shown in Figure \ref{fig:enrichments}.
    %\convmessage{I am looking for a good program to study financial planning that does not require test scores to enroll}
    \item \textbf{Receive a response}:
    Alice's message triggers a response from the AI agent as shown in Turn 1 of Figure~\ref{fig:sample_conv}. The AI agent invokes the two-step process as part of RAG to generate a response: (1) invokes the retrieval model, which then retrieves relevant passages based on Alice's text, (2) invokes the generator model to output a response based on both Alice's text and the retrieved relevant passages.  An example of a response is shown in Figure~\ref{fig:editing-retriever-outputs}.
    %\convmessage{Have you considered Kaplan? They offer a graduate certificate in financial planning. The program is designed to prepare students for the Certified Financial Planner (CFP) certification exam and covers topics like retirement planning, investment and portfolio management, income tax planning, and insurance planning. It is a part-time program that can be completed in 9-12 months. You need a credit score of 620 or higher to enroll. You can also work on becoming a Chartered Financial Analyst (CFA). CFAs work more with institutions and investment firms, focusing on investment analysis and portfolio management. }

    \item \textbf{Review the passages from the retriever model}. This step involves two primary annotator actions (as shown in Figure \ref{fig:editing-retriever-outputs}): 
    \begin{enumerate}
        \item \textbf{Review the attached passages}: \workbench returns at least three relevant passages for each AI agent's generated response. 
        %For each of the contexts, Alice labels them as either relevant or irrelevant via \workbench. 
        Alice reviews the retrieved passages to ensure they are relevant and useful for answering Alice's question. A document is \textbf{relevant} to a turn in the conversation if it is useful to the generator model for generating a response to the human's message. An annotator can mark the retrieved documents from the retrieval model as either relevant or irrelevant. Figure~\ref{fig:sample_conv} shows relevant (green) and irrelevant (red) passages for the two turns.
        %Such annotator feedback is important, especially in cases where the retrieved documents may contain syntactically similar keywords but contain information about semantically different things that might cause the generator model to generate misleading responses. 
        %In Figure \ref{fig:editing-retriever-outputs}, the retriever mistakenly retrieved contexts about credit scores when Alice's message is regarding test scores for enrolling in a course for financial planning. Alice needs to read and review every one of the retrieved contexts. 
        To help her in comparing the passages with the generated AI agent's response, she clicks on the \textbf{overlap icon} in \workbench. The overlap icon highlights the texts that overlap each other in both the passages and the AI agent's response (also shown in Figure \ref{fig:editing-retriever-outputs}). 
        %, as shown in one of the contexts she reviews,
        %\convcontext{While there isn't a single minimum score required by all financial institutions, a \overlaphighlight{credit score of 620 or higher} is generally considered a good starting point of acceptance.} The context is neither relevant nor useful for answering her question, and it is even harmful for the generator model. Alice marks it as irrelevant. 
        
        \item \textbf{Search and attach new passages}: A \textbf{requery} is an annotator action that queries the AI agent's retrieval model for relevant documents based on annotator-specified queries (see Figure \ref{fig:requery} for the requery tool in \workbench). Requerying allows Alice to independently search for relevant documents beyond those that the AI agent initially retrieved. Alice must be cognizant of the primary goal of creating a high-quality conversation as she uses the requery tool: It is important to query with the goal of increasing the coverage of relevant passages for the question, including querying using answer words. Turn 2 in Figure~\ref{fig:sample_conv} has 2 queries. 
        
        %This is especially important in early-version retriever models when the models may not have enough training data to sufficiently retrieve the appropriate set of documents relevant to a particular turn. Alice must be cognizant of the primary goal of creating a high-quality conversation as she uses the requery tool: she must query for contexts that are diverse across all the turns of the conversation. For instance, a conversation is not high-quality if each of the turns contains the same set of contexts. 
    \end{enumerate}
    
    \item \textbf{Update the generator's output} to reflect the latest changes in the attached passages. Alice can click on the regenerate button in \workbench (shown in Figure \ref{fig:editing-retriever-outputs}) to regenerate the AI agent’s response so that it is based on the latest changes in the attached passages.  
    %Alice notices that the updated generator response does not contain the sentence \quotes{You need a credit score of above 620 to enroll.}, as it was erroneously included in the previously generated output due to the context she had just marked as irrelevant.  
    
    \item \textbf{Edit the latest response from the generator model.} An \textbf{edit} is an annotator action that modifies the AI agent's response. Alice edits the AI agent's turn in \workbench (shown in Figure \ref{fig:editing-generator-outputs}). Answer text can be improved by removing (red) or adding (green) text as shown in Figure~\ref{fig:sample_conv}. The AI agent's response must be grounded in the retrieved relevant passages, including the passages that Alice added while using the requery tool. %Alice notices that the latest generated response contains \textbf{inappropriate text} that is irrelevant to her question about financial planning, specifically the ones regarding CFA certifications. She deletes the last two sentences of the generated response:
    %\convmessage{Have you considered Kaplan? They offer a graduate certificate in financial planning. The program is designed to prepare students for the Certified Financial Planner (CFP) certification exam and covers topics like retirement planning, investment and portfolio management, income tax planning, and insurance planning. It is a part-time program that can be completed in 9-12 months. \st{You can also work on becoming a Chartered Financial Analyst (CFA). CFAs work more with institutions and investment firms, focusing on investment analysis and portfolio management. }}
    %- the agent's response needs to be \textbf{faithful}, or grounded, to the relevant contexts. However the response contains unfaithful text about needing a minimum credit score of 670 to enroll in the program. She removes that sentence as well. 
    Alice checks the response for (a) its \textbf{faithfulness}, or groundedness, to the relevant passages, (b) its \textbf{completeness} (the response is not missing any information that can be found in the attached passages), (c) its \textbf{style} (the response is not awkwardly styled), and (d) its \textbf{appropriateness} (the response is useful and concise for answering the question).

    \item \textbf{Enrich the conversation.} \textbf{Enrichments} are custom labels at each turn of the conversation, such as \quotes{Answerable} or \quotes{Unanswerable}, indicating whether a turn is answerable or not (see enrichments in Figure ~\ref{fig:sample_conv} and \ref{fig:enrichments}). Many custom enrichment types can be supported. Alice enriches the turns with task-specific labels, as specified by the instructions given to her. Enrichments are useful for any downstream tasks that ingest the final conversational data. 
    
    \item \textbf{Repeat steps 1 to 5 for the remaining turns.} As Alice moves on to the next turn, \textbf{hints} or tooltips appear at the top of the screen in \workbench whenever there are missing labels, no enrichments, or not enough diverse passages (an example hint is shown in Figure \ref{fig:editing-retriever-outputs}). Alice is done creating all the turns in her conversation when she has met the minimum required length specified in the task instructions. She then exports the conversation in a JSON format. Before she can export the conversation, \workbench will show a checklist, which serves as a final reminder of the properties the conversation needs to meet before exporting (see Figure \ref{fig:export-checklist}). 
\end{enumerate}

The task of creating a conversation cannot be broken up into smaller sub-tasks without compromising the overall quality and natural flow of the conversation~\cite{ragdatahard}. The annotators can "undo" a question if it reduces the flow, passage diversity and/or complexity of the conversation: 1) A question that is immediately answered correctly by the generator and doesn't need any edits can be discarded for lack of complexity; 2) A question whose answer is already part of a prior answer can be discarded for lack of passage diversity; 3) An unanswerable question that follows an unanswerable question can be discarded for poor conversation flow. As a conversation proceeds, the annotator cannot go back and change the questions or passages that occurred in previous turns because it will disrupt the flow of following questions. For quality assurance, we designed a review task, where conversations are reviewed and commented on by other annotators.

\begin{figure}
    \centering
    \begin{subfigure}[t]{0.49\textwidth}
        \centering\includegraphics[width=0.86\columnwidth]{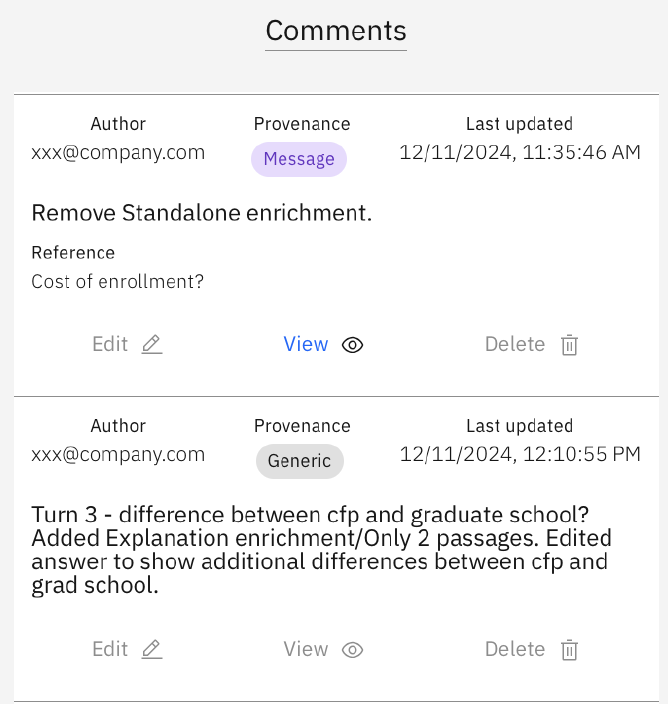} 
    \caption{Annotator comments in \workbench's review mode.}
    \label{fig:review}
    \end{subfigure}
    \centering
    \begin{subfigure}[t]{0.49\textwidth}
        \centering\includegraphics[width=0.94\columnwidth]{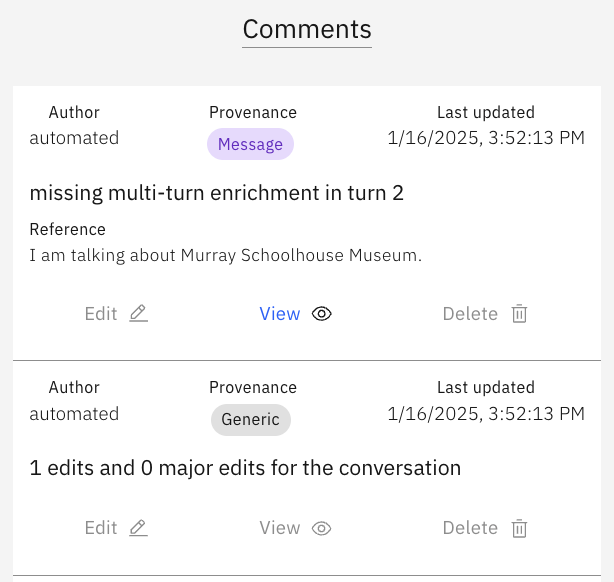}  
    \caption{Automatically generated comments in \workbench's review mode.}
    \label{fig:auto_comments}
        \end{subfigure}
\caption{Examples of annotator and automated comments in review mode.}
\end{figure}

\subsection{Task: Reviewing a conversation for RAG}\label{sec:task-review-conv}

\textbf{Reviewing} a conversation is an annotator task involving adding \textbf{comments} regarding the quality of the conversation and \textbf{editing} the conversation to improve it when possible. Due to the complex nature of the task, all created conversations must go through the review process. In review mode, questions cannot be changed and passages cannot be added as it would disrupt the flow of the conversation. The annotator can only change the relevance of passages, edit the answers, change enrichments and provide feedback via comments. Comments are objects attached to the full conversation or at the turn level to be consumed by annotators as feedback to improve the quality of future conversations. We now walk through how an annotator, Bob, would review the conversation Alice has created, which he views through the review mode in \workbench.

\textbf{Review flow}: Bob reviews the general flow of the conversation, ensuring that it flows naturally and coherently. We explore two scenarios:
    \begin{itemize}
        \item Scenario 1: Bob notices that the conversation talks about enrolling in courses for financial planning, which then naturally moves to the topic of the cost of attending such courses. 
        \item Scenario 2: Bob notices that 
        %the conversation has 3/4 unanswerable questions.
        out of the 4 questions in the conversation, 3 of them are unanswerable.
    \end{itemize}

\textbf{Review complexity}: A conversation is \textbf{complex} if the AI agent finds the annotator's turns challenging to respond to, which results in the AI agent generating a response that may be incomplete or wrong. This then results in the annotator needing to edit the AI agent's response. We quantify \textbf{conversation complexity} through the number of turns that were edited by the annotator. An automated comment is provided to help measure the number of edits (Figure~\ref{fig:auto_comments}). Bob reviews the total number of turns that Alice has edited. The total number of edited turns informs Bob how complex Alice's turns were for the AI agent to answer. A high number of significant edits is desirable for a complex and high-quality conversation. 
\begin{itemize}
    \item Scenario 1: 3 out of 4 turns have significant edits.
    \item Scenario 2: 1 out of 4 turns has a minor edit fixing a grammar mistake.
\end{itemize}

\textbf{Review diversity}: Bob then reviews the passages to ensure \textit{passage diversity}, where each turn should ideally have different (i.e., unique) passages attached. An automated comment regarding the number of unique relevant passages is provided to help measure diversity. Generally, it is allowed to have some passages reappear in other AI agent turns in the same conversation. However, it is desirable for a \textit{high-quality conversation} to have the properties of \textbf{passage diversity}, in which new passages appear at different turns, all the while maintaining the \textit{natural flow of conversation}, i.e. the conversation topic does not seemingly jump between random goals nor do the turns appear disjoint from each other (just to make the passages diverse). 
\begin{itemize}
    \item Scenario 1: 
    %9/15 unique relevant passages in a conversation with 4 turns.
    Out of 15 total passages attached to a conversation with 4 turns, 9 passages were unique.
    \item Scenario 2: 
    %3/4 unique relevant passages in a conversation with 4 turns.
    Out of 4 total passages attached to a conversation with 4 turns, 2 passages were unique.
\end{itemize}

\textbf{Review annotations}: Bob also reviews whether Alice marked the passage relevance correctly and enriched the turns properly. Bob directly fixes any enrichments or labels that were given erroneously. Bob also adds a comment justifying his changes for the task requester and Alice to view at a future time (as shown in Figure \ref{fig:auto_comments}). 
    
\textbf{Accept or reject the conversation}: Bob can accept the conversation if the conversation is \textit{high quality} and \textit{complex} as shown in Scenario 1. Bob can reject the conversation if this is not the case as shown in Scenario 2. As soon as Bob notices that the conversation needs to be rejected he doesn't have to review the full conversation further.

\subsection{Metrics} \label{sec:metrics}

%\sara{Reviewer did not understand the definition of complexity; which we already defined!!! Cite MTRAG paper alongside definitions of conversational complexity. Bring Section 3.3 earlier as the introduction.}

In this section, we describe the metrics from our prior work \cite{katsis2025mtragmultiturnconversationalbenchmark} that we used to measure the quality of the multi-turn conversations from the two annotator populations. We quantify conversation quality based on the \textit{complexity and challenging} level of the conversation: 

\begin{itemize}
        \item \textbf{Average number of turns}: The length of the conversation, computed as the average number of turns in the conversation. Later turns in the conversation tend to be more challenging for the retriever and generator~\cite{katsis2025mtragmultiturnconversationalbenchmark}.
        \item \textbf{Average number of edits}: The average number of turns that have been edited/repaired by the annotator by adding or removing text from the original answer produced by the generator. % We consider this as a measure of complexity for the generator. 
        A conversation that is more complex will have more edits, indicating it was a challenge for the generator.
        \item \textbf{Average number of queries}: The average number of times the annotator queried in the conversation. This includes the initial question and additional re-querying performed per question. We consider this as a measure of complexity for the retriever. A conversation that is more complex will need more queries to find more relevant passages.
        \item \textbf{Average number of unique passages}: The average number of unique relevant passages for all turns in the conversation. We consider this as a measure of passage diversity. A conversation is not diverse if it has few unique passages and a conversation is diverse if it has many unique passages. A conversation that is more complex will be more diverse. 
    \end{itemize}
The higher the value of all of these metrics, the more likely it is considered a \textit{complex and challenging conversation} which makes it a higher quality conversation for evaluating retrievers and generators. These metrics are a proxy for quality though it may sometimes not be the case, particularly along one dimension; e.g. a conversation can have many turns and still be of poor quality. In Section~\ref{sec:datacollection} we show that these metrics align with the human understanding of conversation quality based on the accept/reject rate of conversations. Further details about these metrics and their motivation can be found in ~\cite{katsis2025mtragmultiturnconversationalbenchmark} .

%% file: tbl/survey_demographics.tex
\begin{table}[tb]
    \small
    \centering
    %{p{0.6\columnwidth}p{0.17\columnwidth}p{0.05\columnwidth}}%
    \begin{tabular}{ll|l|l|l|l|ll|l}
        \toprule
            Group 
            & PID   
            & \makecell[l]{Annotation\\Experience} 
            & \makecell[l]{Technology\\Familiarity} 
            & \makecell[l]{AI\\Background}  
            & \makecell[l]{RAG\\Background} 
            & Age (Years)   
            & Gender 
            & \makecell[l]{\# Created\\Conv}    \\
        \midrule
        \midrule
            \multirow{24}{*}{Ext }  
            & Ext1  
            & Less than 1 year
            & \veryfamiliar      
            & \competent         
            & \advancedbeginner 
            & 50 to 60        
            & Female 
            & 75 to 100 
        \\
            & Ext2  
            & 1 to 3 years                              
            & \slightlyfamiliar  
            & \beginner          
            & \beginner          
            & 40 to 50        
            & Female 
            & 100+
        \\
            & Ext3  
            & 1 to 3 years
            & \slightlyfamiliar  
            & \proficient        
            & \expert            
            & 18 to 30        
            & Female 
            & 100+
        \\
            & Ext4  
            & 1 to 3 years
            & \somewhatfamiliar  
            & \beginner          
            & \advancedbeginner 
            & 40 to 50        
            & Female 
            & 100+ 
        \\
            & Ext5  
            & 1 to 3 years
            & \somewhatfamiliar  
            & \advancedbeginner 
            & \beginner          
            & 18 to 30        
            & Female 
            & Less than 25
        \\
            & Ext6  
            & 1 to 3 years                              
            & \somewhatfamiliar  
            & \advancedbeginner 
            & \beginner          
            & 30 to 40        
            & Female 
            & 25 to 50 
        \\
            & Ext7  
            & 1 to 3 years                              
            & \somewhatfamiliar  
            & \advancedbeginner 
            & \beginner          
            & 40 to 50        
            & Male   
            & 100+ 
        \\
            & Ext8  
            & 1 to 3 years                              
            & \somewhatfamiliar  
            & \advancedbeginner 
            & \beginner          
            & 40 to 50        
            & Male   
            & 50 to 75
        \\
            & Ext9  
            & 1 to 3 years                              
            & \somewhatfamiliar  
            & \advancedbeginner 
            & \competent         
            & 30 to 40        
            & Female 
            & 75 to 100
        \\
            & Ext10 
            & 1 to 3 years                              
            & \somewhatfamiliar  
            & \competent         
            & \competent         
            & 30 to 40        
            & Female 
            & 75 to 100
        \\
            & Ext11 
            & 1 to 3 years                              
            & \veryfamiliar      
            & \competent         
            & \competent         
            & 30 to 40        
            & Male   
            & 50 to 75
        \\
            & Ext12 
            & 1 to 3 years                              
            & \veryfamiliar      
            & \proficient        
            & \proficient        
            & 30 to 40        
            & Female 
            & 100+
        \\
            & Ext13 
            & 1 to 3 years                              
            & \veryfamiliar      
            & \proficient        
            & \proficient        
            & 50 to 60        
            & Female 
            & 100+
        \\
            & Ext14 
            & 1 to 3 years
            & \extremelyfamiliar 
            & \proficient        
            & \beginner          
            & 40 to 50        
            & Male   
            & 75 to 100
        \\
            & Ext15 
            & 1 to 3 years                            
            & \extremelyfamiliar 
            & \proficient        
            & \proficient        
            & 30 to 40        
            & Female 
            & 100+
        \\
            & Ext16 
            & 1 to 3 years                            
            & \extremelyfamiliar 
            & \proficient        
            & \proficient        
            & 50 to 60        
            & Female 
            & 25 to 50
        \\
            & Ext17 
            & 3 to 5 years                            
            & \veryfamiliar      
            & \competent         
            & \competent         
            & 30 to 40        
            & Female 
            & 100+
        \\
            & Ext18 
            & 3 to 5 years 
            & \veryfamiliar      
            & \proficient        
            & \competent         
            & 50 to 60        
            & Male   
            & 75 to 100
        \\
            & Ext19 
            & 3 to 5 years                            
            & \extremelyfamiliar 
            & \expert            
            & \competent        
            & 18 to 30        
            & Female 
            & 100+
        \\
            & Ext20 
            & 3 to 5 years                            
            & \extremelyfamiliar 
            & \expert            
            & \expert            
            & 18 to 30        
            & Male   
            & 25 to 50
        \\
            & Ext21 
            & 5 to 10 years
            & \extremelyfamiliar 
            & \expert            
            & \proficient        
            & 18 to 30        
            & Male   
            & 25 to 50
        \\
            & Ext22 
            & 10+ years                               
            & \somewhatfamiliar  
            & \proficient        
            & \competent         
            & 30 to 40        
            & Male   
            & 50 to 75
        \\
            & Ext23 
            & 10+ years                               
            & \extremelyfamiliar 
            & \competent         
            & \competent         
            & 40 to 50        
            & Female 
            & 100+
        \\
            & Ext24 
            & 10+ years                               
            & \extremelyfamiliar 
            & \expert            
            & \expert            
            & 30 to 40        
            & Male   
            & 50 to 75
        \\
        \midrule
        \midrule
            \multirow{7}{*}{Int }    
            & Int1  
            & 1 to 3 years                            
            & \extremelyfamiliar 
            & \beginner          
            & \beginner          
            & 60+             
            & Female 
            & 25 to 50
        \\
           & Int2  
           & 1 to 3 years                               
           & \extremelyfamiliar 
           & \advancedbeginner 
           & \advancedbeginner 
           & 18 to 30        
           & Male   
           & 100+
       \\
           & Int3  
           & 5 to 10 years                            
           & \veryfamiliar      
           & \beginner          
           & \advancedbeginner 
           & 60+             
           & Female 
           & 25 to 50
       \\
           & Int4  
           & 5 to 10 years                           
           & \veryfamiliar      
           & \competent         
           & \advancedbeginner 
           & 60+             
           & Female 
           & 75 to 100
       \\
           & Int5  
           & 10+ years                               
           & \somewhatfamiliar  
           & \beginner          
           & \beginner          
           & 60+             
           & Female 
           & 100+
       \\
           & Int6  
           & 10+ years                                
           & \somewhatfamiliar  
           & \beginner          
           & \beginner          
           & 60+             
           & Female 
           & 100+
       \\
           & Int7  
           & 10+ years                                  
           & \veryfamiliar      
           & \competent         
           & \competent  
           & 60+             
           & Female 
           & 75 to 100
       \\      
        \bottomrule
    \end{tabular}
    \caption{Demographics of all 7 internal annotators (IntAnno) and 24 external annotators (ExtAnno). Each bar represents a rating on the Likert scale, i.e. 1 bar represents \quotes{Beginner} and 5 bars represent \quotes{Expert}. \# Created Conv denotes the reported number of conversations the annotator has created at the time of collecting the demographics after 1 3/4 years of the task.
    %(which was after all three tasks were given to them).
    }
    \label{tbl:demographics}
\end{table}

%% file: method.tex
\section{Method}
\label{sec:method}

The task of creating multi-turn conversations was carried out during an approximately \studyduration using the two annotator populations of different feedback loops: internal and external. The populations consisted of 7 internal annotators and 40 external annotators. In this section, we describe the annotator's communication structures and demographics (Section \ref{sec:annotators}) and the method for data collection during the longitudinal study (Section \ref{sec:datacollection}).

\subsection{Annotators} \label{sec:annotators}

Our study explored two groups of annotators: 7 internal and 40 external annotators. Internal annotators were from the same company as the researchers of this paper who created and defined the task. External annotators were recruited through an external, professional annotation company. %\footnote{The findings from this paper are not dependent on the external annotation service we used.}. 

\subsubsection{Communication feedback structures and material} 

We did not have the ability to directly communicate with the external annotators. All communication was done via an intermediary represented by the external annotation service. Task instructions and materials, such as tutorials, including videos and slide decks, were sent to external annotators through the intermediary. We met with the intermediary weekly to discuss progress and share feedback. The annotators were able to send us questions through the intermediary but they could not ask us directly. Responses and feedback were returned in a batch to the intermediary who could then choose how to disseminate the feedback to the annotators. In contrast, we communicated directly with the internal annotators through e-mail, instant messaging in Slack groups or direct messages and video calls on Microsoft Teams and shared folders for all material. The annotators were able to ask questions and receive immediate and personal replies and feedback. We met with the internal annotators individually and in a group frequently, as needed. 

% \subsubsection{Instruction materials} 
% Both annotation populations were provided with slides with instructions. The internal annotators were also provided with a live training session which included going over the instructions, how to use the tool and create a conversation. In addition, a Slack Channel was setup for questions and discussions for direct and immediate feedback. The external annotators were provided with written guidelines that include instructions and a video showing how to use the tool and create conversations. We did not have direct access to the external annotators; there was an intermediary for any communication and questions.  

%\todo{Add something about weekly meeting, external annotators met weekly with the intermediary}

\subsubsection{Demographics \& Skill Level} 

We collected detailed information about the annotators demographics and \skilllevel after all three primary tasks were given to both groups of annotators (see Table \ref{tbl:demographics}). 
We use the term \textit{\skilllevel} to refer to the annotator's skill level in the RAG conversational creation task. It is important to note that while an annotator may have created a lot of conversations, it does not necessarily translate to them being good/skilled at the task (hence the term \skilllevel). 

Given the nature of our communication structure with the external annotators, only a subset of the 40 external annotators who participated in the data creation tasks also chose to complete the survey. We asked the annotators their gender, age, years of professional annotation experience, and additional questions regarding their technical background (rated over a Likert scale from 1 to 5):
    \begin{itemize}
        \item Technology Familiarity: \quotes{On a scale of 1-5, would you say you are familiar with most common computer programs and online tools?}, with 1 being \quotes{Not familiar at all} and 5 being \quotes{Extremely familiar}.
        \item AI Background: \quotes{On a scale of 1-5, with 1 being beginner and 5 being expert, how would you rate your understanding of AI?}
        \item RAG Background: \quotes{On a scale of 1-5, with 1 being beginner and 5 being expert, how would you rate your understanding of RAG?}
    \end{itemize}

All our professional annotators exhibited the same diversity that previous research~\cite{wang2024case, agreementnotgold2023, asher-etal-2016-discourse, ensemble2014, shepherding2012} describes. Compared to the internal annotators, more external annotators report having a competent (3 on a Likert scale) understanding of AI and RAG. The vast majority of the external annotators (16/24) report having less than 3 years of annotation experience, while a majority of internal annotators (5/7) have at least 5 years of annotation experience. Our internal annotators generally have more experience in annotation and are much older (60+ years old) than the external annotators. 

\subsubsection{Compensation} 
All annotators from both groups are considered professional annotators and were compensated well above minimum wage for their annotation expertise. Internal annotators were paid hourly and external annotators were paid per accepted conversation (low-quality conversations may lead to a rejection or a redo of the task without additional pay). The pay structure, which is an important aspect of the feedback loop, may have impacted the quality of the work, as previous studies have shown that differences in financial incentives impact participant recruitment \cite{financialincentives2014, hcicompensation2021}. %\todo{cite some paper here}%\todo{cite Incentives to Participate in Online Research: An Experimental Examination of “Surprise” Incentives  } 

\subsection{Phases of Data Collection} \label{sec:datacollection}

\input{tbl/phase_data}

The total number of conversations created was approximately 1500 by the internal annotators and 5000 by the external annotators. In this section, we describe and report results of our longitudinal study for both annotator populations on a subset of the data for the three phases of the task: (1) pilot phase, (2) creation phase, and (3) review phase. Our findings are shown in Tables~\ref{tab:comparison} and ~\ref{tab:review}. We also explore synthetic data as an alternative to human annotations (Section \ref{sec:synthetic}).

% \begin{enumerate}
    % \item Pilot phase (May to August 2024)
    % \item Creation phase (Sept 2024 to Present)
    % %\item Synthetic phase (Sept to October 2024)
    % \item Review phase (November 2024 to Present) 
% \end{enumerate}

\subsubsection{Phase: Pilot}

\begin{figure*}
    \centering
    \includegraphics[width=\linewidth]{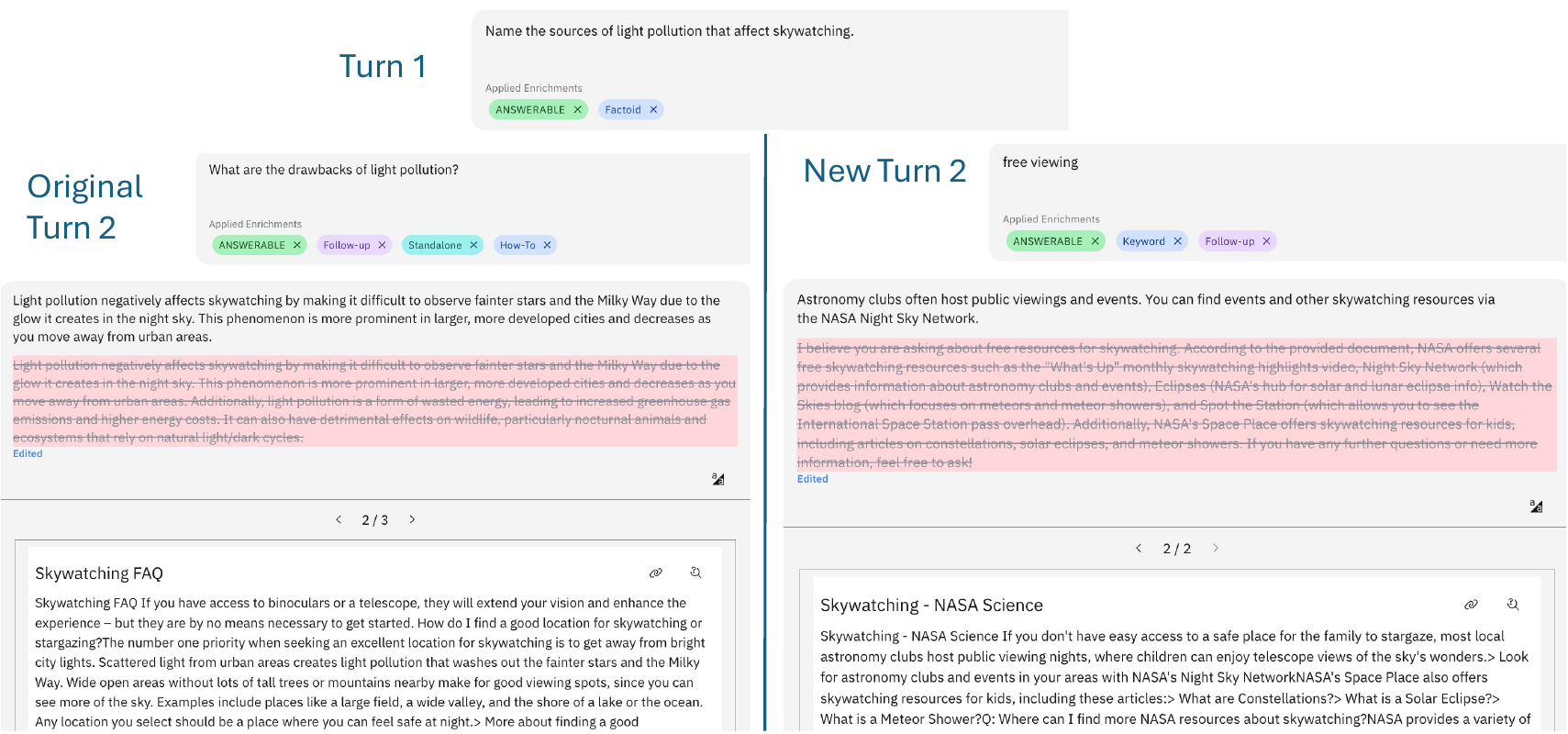}
    \caption{Comparison of original turn and more complex alternative. The original turn 2 was not complex. It provided the same three passages and repeated a lot of the same information in the answer for Turn 1. The New Turn 2: 1) is more ambiguous because it is a keyword, 2) it requires a significant amount of editing, and 3) it required re-querying which was used to find a new relevant passage as shown.
}
    \label{fig:complexity}
\end{figure*}

The pilot task was a small-scale experiment to measure whether the task and instructions were clear and to calibrate the task as needed. The statistics from the pilot are shown in Table~\ref{tab:comparison}. In this task we reviewed the conversations from both annotator populations to gain insights into how the annotators performed the task. The initial feedback was the same, the conversations were not complex enough and too easy for the agent. There was on average one edited answer and only one unique passage per question for the entire conversation. Our feedback was to ask questions that diversify the relevant passages and that the questions should be challenging enough that the initial answer provided by the agent should need to be edited. A screenshot comparing what a question looked like in the pilot changed to a more complex question is shown in Figure~\ref{fig:complexity}. %\todo{describe the kinds of instruction materials we gave to the different populations, e.g. slides, videos, what kinds of content they had (at a high-level)}

\subsubsection{Phase: Creation}
\label{sec:synthetic}

%Following the pilot, we reviewed the conversations to gain insights into how the annotators performed the task. We identified common mistakes and areas of improvement, particularly the need for the conversations to be more complex and diverse. 
Following the review of the pilot we shared feedback with both groups of annotators by updating the slides to improve clarity and describe the desired complexity. We also provided several examples showing how to improve existing conversations. We had virtual meetings with the internal annotators and updated the additional instructions provided to the external annotators. Further, we added additional features to the tool to improve the tooling and encourage more complex conversations. We added hints containing tips and an export checklist as shown in Figures~\ref{fig:editing-retriever-outputs} and ~\ref{fig:export-checklist}. The annotators then began creating new conversations that were more complex.

A small pilot was provided with some real-time and follow-up feedback to the internal annotators and then the full task began. A comparison of conversations created during the complexity task is shown in Table~\ref{tab:comparison}. The number of conversations for the external annotators is following a review by the QA team of the external provider where 309/800 conversations were rejected. Since the external annotators are a larger population group they can create considerably more conversations. However, the internal annotator conversations are longer, have more edits, and considerably more querying and unique passages than the external annotator conversations.

% \todo{From yannis: need a better transition here, since the synthetic conversations appear out of the blue and it is not fully clear how they connect to the rest of the work.} 
An LLM can also be used to generate conversations as an alternative to human-generated conversations. The advantage is clear: an LLM can generate significantly more conversations in a very short span of time. 
% \todo{From yannis: it is not clear why MTRAG was selected and how it compares to our conversations in terms of underlying corpora. An alternative would be to say that we experimented with creating synthetic conversations over the same corpora leveraging state-of-the-art techniques (and use the same citations as in the synthetic data section of the MTRAG paper). } 
We leverage state-of-the-art techniques ~\cite{lee2024multidocumentgroundedmultiturnsynthetic} to generate synthetic conversations over our corpora and compare them to the manually created conversations in Table~\ref{tab:comparison}. 
% We analyze the synthetic conversations from MTRAG~\cite{lee2024multidocumentgroundedmultiturnsynthetic,katsis2025mtragmultiturnconversationalbenchmark} and compare them to the manually created conversations in Table~\ref{tab:comparison}. 
The synthetic conversations are of reasonable length, but not as long as the conversations of the internal annotators as the conversation tends to degrade as it gets longer~\cite{lee2024multidocumentgroundedmultiturnsynthetic}. These conversations also have less passage diversity than the conversations from both human populations. There are no edits or requerying by the synthetic data generator so the quality may be lower which we will analyze in the next section.

\subsubsection{Phase: Review}

Following the completion of the task of creating conversations, a review phase was performed. All conversations went through an automated review where comments were generated automatically to address passage diversity, amount of edits, and missing enrichments. A conversation is considered to have passage diversity if for the amount of Questions, $Q_n$, the number of unique relevant passages, $P$, is $P >= (Q_n-1) \times 2$. So a conversation with 3 questions should have at least 4 passages and a conversation with 5 questions should have at least 8 passages. We identified both major and minor edits by computing the Rouge-L score using HuggingFace\footnote{\url{https://huggingface.co/docs/evaluate/en/choosing_a_metric}} between the original and edited response. Major edits were more important than minor changes (e.g. typo fixes). An example of the automated comments is shown in Figure~\ref{fig:auto_comments}. These comments were used to help guide the annotators during manual review.

Next, the internal annotators manually reviewed the conversations from both annotator pools and synthetic conversations. In some cases, this was a subset of the annotations created. The review provides the ability to accept, reject or edit and accept the conversations and  provide comments as feedback. 87\% internal, 69\% external and 72\% of the synthetic conversations were accepted. The repair during the review increased the number of edits and decreased the number of relevant passages for all populations. Even though a nice amount of the synthetic conversations were accepted they were considerably less diverse than the human conversations as passages can not be added during the review phase. The accepted conversations created and then reviewed by the internal annotators during this phase were used in our released MTRAG~\cite{katsis2025mtragmultiturnconversationalbenchmark} benchmark. All other accepted conversations can be used as future training data. 
%\todo{Sara to add the correct citation here, cannot use REDACTED citations as paper will desk rejected}

% \subsection{Comparisons with Synthetic}

%\subsection{Analysis}

\subsection{Survey}

\begin{figure}
    \centering
    \begin{subfigure}[t]{0.43\textwidth}
        \centering
        \includegraphics[width=\linewidth]{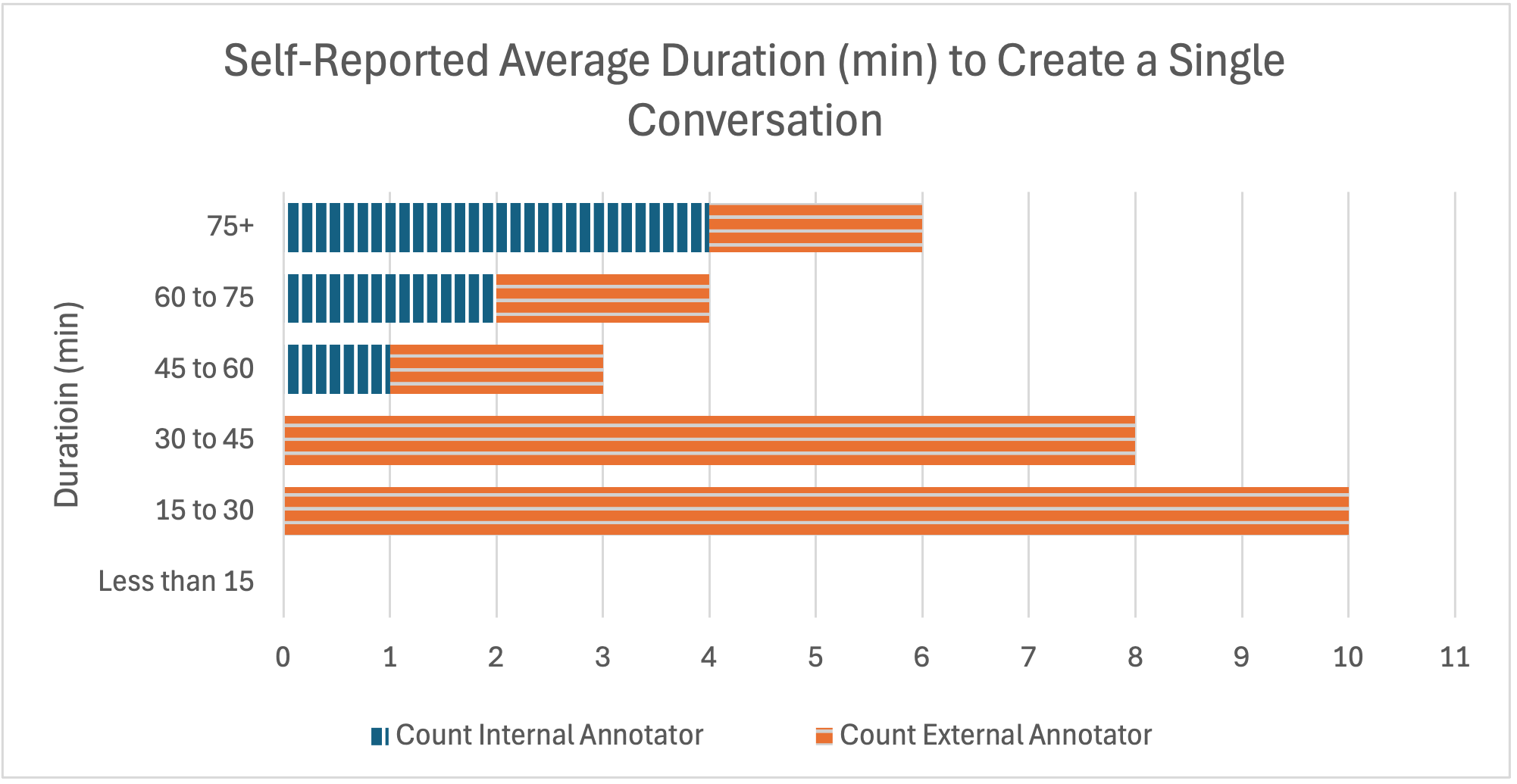}
        \caption{Self-reported average time it takes for annotators to create a single conversation.}
        \label{fig:duration_create_conversation}
    \end{subfigure}
    \begin{subfigure}[t]{0.43\textwidth}
        \centering
        \includegraphics[width=\linewidth]{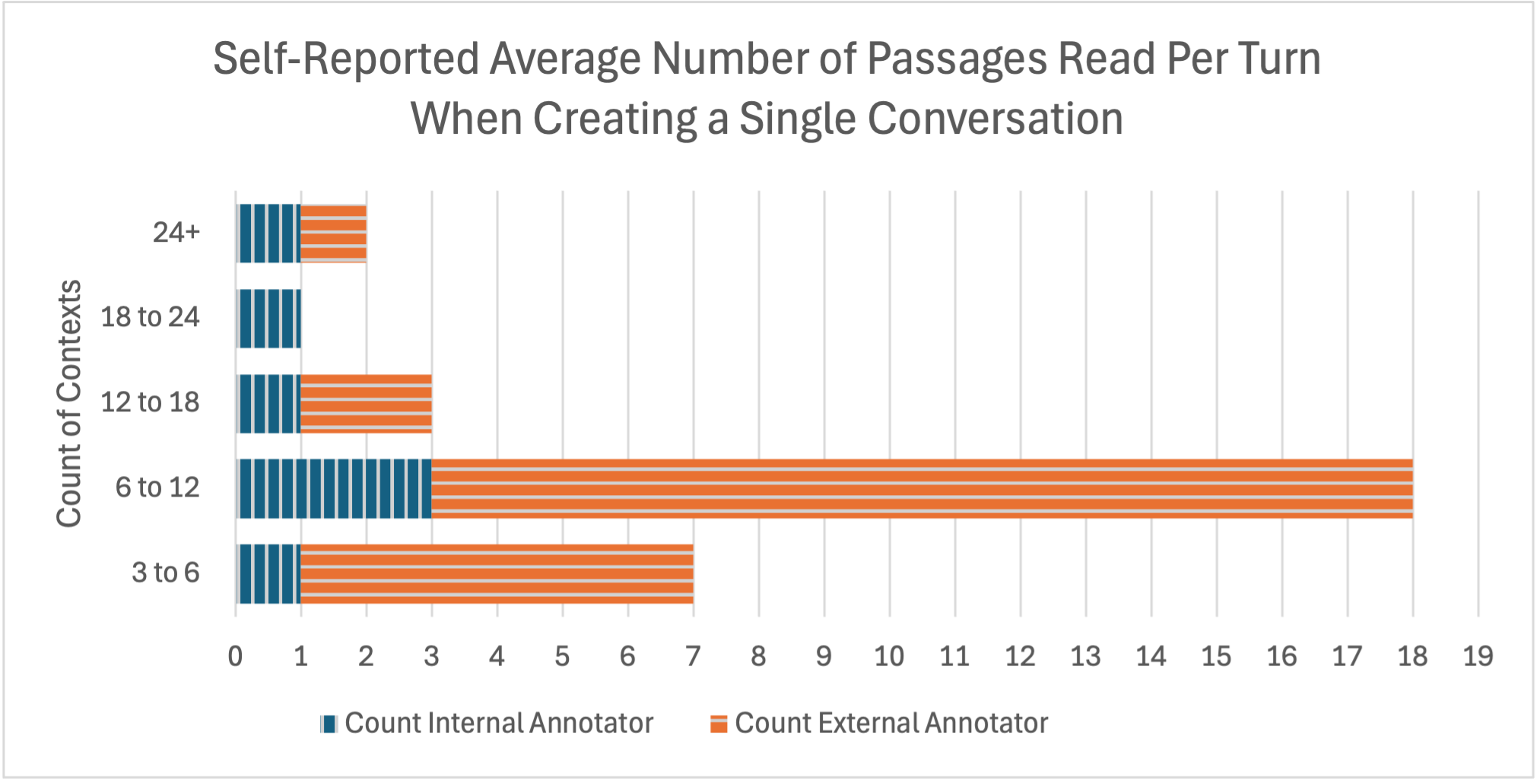}
        \caption{Self-reported average number of passages annotators read per turn when creating conversations.}
        \label{fig:context_create_conversation}
    \end{subfigure}
    \caption{Survey results illustrate the difficulty and time it takes to create a single RAG conversation.}
    \label{fig:survey-task-hardness}
\end{figure}

\begin{figure}
    \centering
    \begin{subfigure}[t]{0.49\textwidth}
        \centering
        \includegraphics[width=0.95\linewidth]{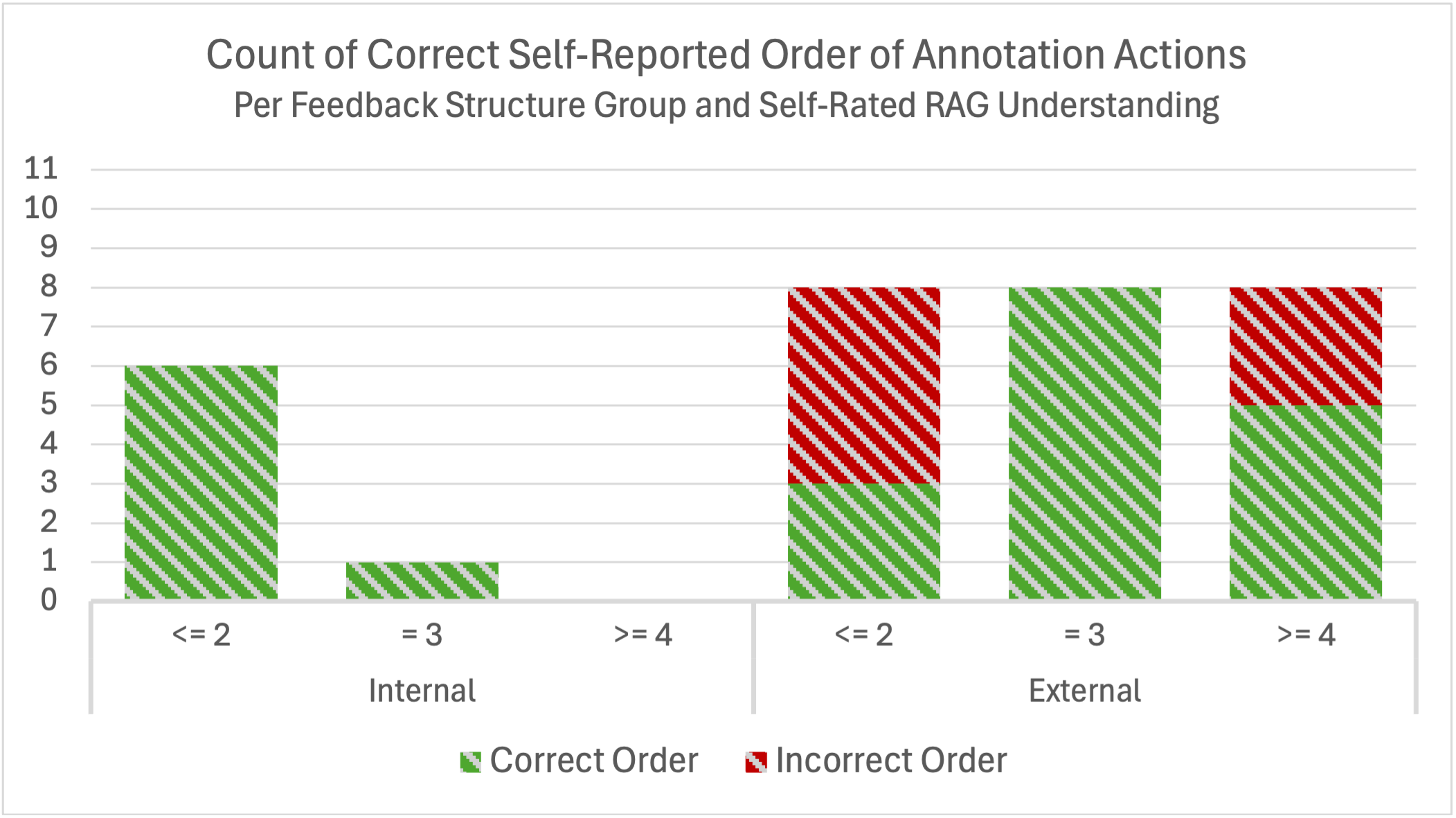}
        \caption{Self-reported understanding of RAG (on a Likert scale of 1 \quotes{Beginner} to 5 \quotes{Expert}).}
        \label{fig:annotator_actions_order_result}
    \end{subfigure}
    \hspace{1pt}
    \begin{subfigure}[t]{0.49\textwidth}
        \centering
        \includegraphics[width=0.95\linewidth]{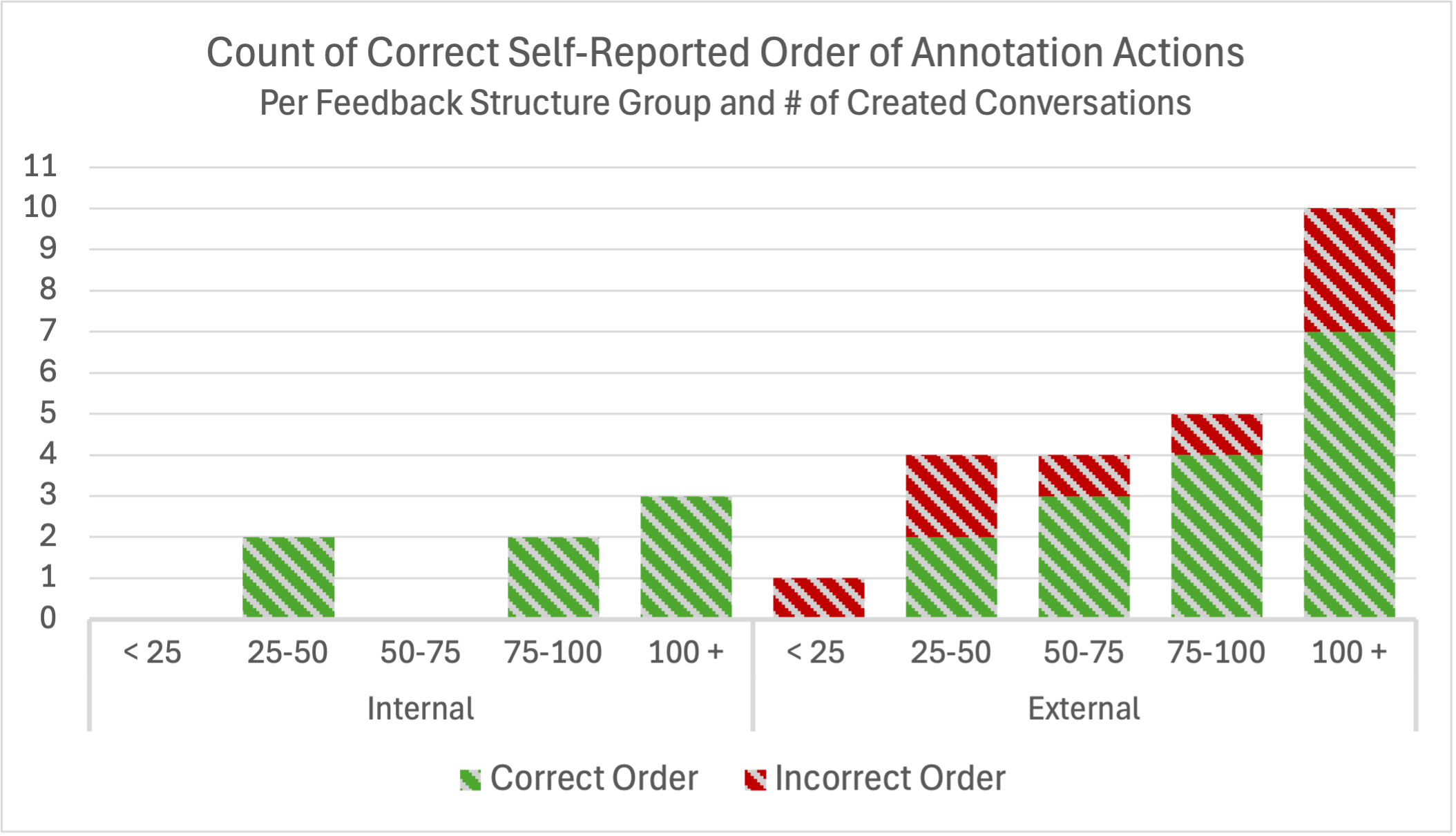}
        \caption{Self-reported number of conversations created.}
        \label{fig:annotator_actions_order_grouped_skill}
    \end{subfigure}
    \caption{Order of annotator actions per turn when creating a conversation. The correct order of actions are as follows: (A) Create a turn, (B) Edit the outputs of the retriever model, and (C) Edit the latest response from the generator model. Counts are grouped based on the feedback structure (internal/external).}
    \label{fig:annotator_actions_order}
\end{figure}

% Link to survey results are here: https://ibm.ent.box.com/file/1814509511960

We surveyed both groups of annotators in February 2025, after all three major tasks had been given to them. We wanted to understand how the different feedback structures impacted the quality of the created data. We also wanted to understand how the different feedback structures impacted how they performed and perceived the task.  

To quantify their skill level in the complex task of creating RAG conversations  after the three tasks: we asked them, \quotes{Approximately, how many conversations have you created using Workbench?} (see Table \ref{tbl:demographics}). 15/24 external annotators and 5/7 internal annotators reported that they created at least 75 conversations. 

\subsubsection{Feedback structure impacts the duration and effort to complete the task} 
% \todo{From yannis: The financial incentive structure may have also played a role in this (with the external annotators being paid per conversation, whilte internal annotators were paid hourly). }
To understand how the different feedback structures impacted the duration and effort it took for them to create a single conversation, we asked annotators:
    \begin{itemize}
        \item \quotes{Approximately, how long (how many minutes) does it take to create a single conversation?} Analysis of the time it takes to do the task is shown in Figure \ref{fig:duration_create_conversation}.
        \item \quotes{Approximately, how many passages do you feel you read at each turn when you create a conversation?} Analysis of the average number of passages annotators read per turn when creating a single conversation in \workbench is shown in Figure \ref{fig:context_create_conversation}. 
    \end{itemize}

Most of the external annotators reported that it took them less than 45 minutes to create a single conversation, while all of the internal annotators reported that the task took at \textit{least 45 minutes} to complete. The internal annotators remarked on the time it took to read the passages at each turn, as Int4 says, \quotes{Way too much reading and re-reading is involved.} and Int3 says, \quotes{It takes time to carefully read.} Interestingly, no external annotators have expressed such concerns in the survey. While the instruction materials emphasized reading and verifying all of the attached passages, the distant feedback structure between the task requester and the external annotators influenced the reported speed and care the external annotators took to deal with core sub-tasks in creating a RAG conversation. 

We also observe a similar disparity in their \textit{effort}, or self-reported number of passages that are read per turn when creating a conversation (see Figure \ref{fig:context_create_conversation}). External annotators, on average, read fewer passages per turn than the internal annotators. Most of the external (15/24) and internal annotators (3/7) reported to have read an average of 6 to 12 passages per turn. We emphasize in the instructions that they need to read all of the attached passages, 3 of which are the default number of passages that \workbench attaches to the response after the annotators create a turn. They are also required to requery and read (or at a minimum skim) the additional passages from requerying. Each requery returns 9 passages ranked by the retriever for relevance to the requery words that were used. Hence, the total read passages at a turn should be around 12 if an annotator indeed read all 3 attached passages in the response and passages from several requeries. 
% There was a subset of the external annotators (6/24) and a single internal annotator who reported a range under the recommended number of contexts to read per turn. \todo{say something about that internal annotator who reported reading a number of contexts lower than the recommended range} 

\subsubsection{Feedback structure impacts the order of annotator actions in the task}

We wanted to understand the impact of the different feedback structures on their understanding of how one should create a turn in a RAG conversation, specifically the order of annotator actions in a turn when creating a conversation. We asked annotators to order the actions they perform when they create a turn in a conversation.  Ideally, the following should be the order of actions: (A) Create a question, (B) Edit the outputs of the retriever model (which includes verifying the retrieved passages, requerying, and labeling the attached passages), (C) Edit the latest response from the generator model (which includes regenerating the agent's response and editing it). Enriching the turn can occur at any point in time. See Figure \ref{fig:annotator_actions_order}, where we grouped the counts based on the feedback structure (internal/external) and their self-rated understanding of RAG (on a Likert scale of 1 to 5, with 1 being \quotes{Beginner} and 5 being \quotes{Expert}). 

We observe that none of the internal annotators reported an understanding of RAG greater than 4 (\quotes{Proficient} and \quotes{Expert}) while all reported the expected order of annotator actions. On the other hand, some external annotators reported a more proficient understanding of RAG while also reporting an incorrect order of annotator actions when creating a turn in a RAG conversation. Some of the external annotators also have reported a technical background with a proficient background in AI (see Table \ref{tbl:demographics}), which we confirm as either formal training in school or courses in AI. However, even with such a background, it is clear how the impact of a different feedback loop affects the annotator's performance in completing complex tasks. 

%There were two internal annotators with the wrong order of annotator actions (reporting that the regeneration of the agent's answer should be done towards the end, i.e., ABDC). This is not linked to the impact of the communication feedback structure with the task requester. Rather, it is attributed to their perception of the generator model, which was an earlier version that would often not drastically change the generated output when there were changes in the attached passages. However, they understood that the outputs needed to be regenerated when the passages changed. Participant, Int4, says, \quotes{The agent['s] response is often very good, but not supported by documents. Therefore, we need to edit it.} %\item Participant, Int4, says, \quotes{I edit the conversation to make it faithful, not the tool.}[Int4]  them} 

We also charted the number of annotators with correct and incorrect order of annotator actions, and grouped those counts based on their feedback structure and their skill level in creating RAG conversations (i.e., the number of conversations they created) (see Figure \ref{fig:annotator_actions_order_grouped_skill}). 14/24 external annotators created more than 50 conversations and reported the correct order of annotator actions, indicating that skill level and practice have also helped them understand how to correctly complete the complex task. However, for some of the external annotators, a higher skill level does not translate to reporting the correct, expected order of annotator actions; 10/24 external annotators created more than 100+ conversations, but 3 of them did not report the correct order of annotator actions. 

\subsubsection{Feedback structure impacts annotator perceptions of the influence the tool features have on data quality}\label{sec:annotator-perception-tool-feature}

We also asked the annotators regarding their perception of different tool features and their impact on the quality of the conversation if we were to remove that feature from \workbench. Specifically, we asked them, \quotes{On a scale of 1 to 5, if we were to remove the ability for you to use <INSERT \workbench FEATURE>, how would this decrease the quality of the conversations you create?}, where 1 is \quotes{No decrease in quality at all}  and 5 is \quotes{Extreme decrease in quality}. We computed the absolute mean difference ($\mu$ difference) between the two populations (see Table \ref{tbl:survey}).

There were larger differences between internal and external annotators on the tool's hints ($\mu$ difference = 1.41), which are pop-ups that appear as the annotator creates the RAG conversation (see Figure \ref{fig:editing-retriever-outputs}). Hints appear on demand when the tool detects issues in the data quality, such as missing annotations or low counts of diverse passages with respect to the number of turns in the current conversation. Intuitively, removing hints from the tool would make the already-complex task even harder, especially when one group of annotators has an indirect feedback loop with the task requesters.

There were also larger differences between internal and external annotators regarding the tool features for changing the retrieved passages, specifically requerying ($\mu$ difference = 0.78), marking passages relevant/irrelevant ($\mu$ difference = 0.78), and the yellow highlights that mark texts overlap in the passages and the generated response ($\mu$ difference = 0.75). The external annotators did not view changing the passages as important as the internal annotators. This is also reflected in the disparity between the self-reported number of average passages the external annotators read per-turn (see Figure \ref{fig:context_create_conversation}) as well as the average number of unique passages that were actually marked as relevant (see Table \ref{tab:comparison}), suggesting that the disjointed feedback structure made them not notice the importance of editing the retriever model outputs at the same level of consciousness as the internal annotators did. 
%It is important to emphasize that both groups of annotators used the tool the same amount of time, yet the different feedback loop significantly impacted the quality of the final data. 

On the other hand, editing the generator model's outputs is viewed equally important by both internal and external annotators ($\mu$ difference = 0.04). %This may be due to the shared perception that the generator model (an early version they were using for the tasks) did not generate high-quality responses. 
An external annotator, Ext24, says, \quotes{Without this functionality, it would often be impossible to generate conversations according to the instructions.} An internal annotator, Int2, also echoes the same sentiment, \quotes{Sometimes the initial agent answer has too much additional/unnecessary information.} 

%% file: tbl/phase_data.tex
\begin{table}[]
    \centering
    \begin{tabular}{l|rr|rr|r}
    \toprule
\multirow{2}{*}{Metrics } & \multicolumn{2}{c|}{Pilot} &	\multicolumn{2}{c|}{Creation} & Synthetic \\
\cmidrule{2-6}
&	External &	Internal &	External	& Internal	& Pre-Review \\
\midrule
\midrule
    Num conversations &	33 &	53 &	\textbf{491} &	127 &	200 \\
\midrule
Avg num turns  &	4.0 &	3.7 &	4.2 &	\textbf{7.6} &	5.9 \\
Avg num edits &	1.1 &	1.5 &	3.0 &	\textbf{7.0} &	- \\
Avg num queries &	1.0 &	1.0 &	6.2 &	\textbf{12.7} &	NA \\
Avg num unique passages &	4.5 &	4.0 &	7.3 &	\textbf{17.1} &	4.6 \\
\bottomrule
    \end{tabular}
    \caption{Comparison of conversations per task and annotator group for the Pilot and Create phases}
    \label{tab:comparison}
\end{table}

\begin{table}[]
    \centering
    \begin{tabular}{l|rr|rr|rr}
    \toprule
    \multirow{2}{*}{Metrics }
& \multicolumn{2}{c|}{External} &	\multicolumn{2}{c|}{Internal} & \multicolumn{2}{c}{Synthetic} \\
\cmidrule{2-7}
&	Pre &	Post &	Pre	& Post	& Pre & Post \\
    \midrule
    \midrule
Num conversations &	251 &	173 &	127 &	110 &	75 &	53 \\
\midrule
Avg num edits &	2.9 &	3.4 &	7.0 &	7.3 &	- &	2.0 \\
Relevant yes  &	10.1  &	8.4 &	20.0 &	19.4 &	- &	11.2 \\
Relevant no	 & 5.4 &	5.2 &	8.7 &	9.4 &	-  &	13.9 \\
\bottomrule

    \end{tabular}
    \caption{Comparison of the statistics pre- and post- Review Phase for the main task}
    \label{tab:review}
\end{table}

%% file: tbl/survey_internal_anno.tex
\begin{table}[tb]
    \small
    \centering
    \begin{tabular}{l|rr|rr|r}%{p{0.6\columnwidth}p{0.17\columnwidth}p{0.05\columnwidth}}%
        \toprule
            \multirow{2}{*}{Tool Feature }
                & \multicolumn{2}{c|}{IntAnno} 
                & \multicolumn{2}{c|}{ExtAnno } 
                & \multirow{2}{*}{\makecell[c]{$\mu$\\Difference}}\\
            \cmidrule(lr){2-3}\cmidrule(lr){4-5}
                & $\mu$
                & $\sigma$ 
                & $\mu$
                & $\sigma$ \\
        \midrule
        \midrule
            Hints 
            & \textbf{4.29}   
            & 0.76
            & 2.88	
            & 1.42 
            & 1.41\\
        \midrule
            Requery Tool 
            & \textbf{4.57}
            & 0.53
            & 3.79
            & 1.50 
            & 0.78\\
        \midrule
            Marking passages relevant/irrelevant
            & \textbf{3.86}
            & 1.07 
            & 3.08
            & 1.50 
            & 0.78\\
        \midrule
            The overlap icon highlights overlapping text in the passage and response 
            & \textbf{4.71} 
            & 0.49
            & 3.96	
            & 0.91 
            & 0.75 \\
        \midrule
            Regenerating the agent response
            & 3.86
            & 0.90
            & \textbf{4.17}	
            & 1.01 
            & 0.31 \\
        \midrule
            Enriching the questions
            & 2.57
            & 1.33
            & \textbf{2.79}
            & 1.28 
            & 0.22\\
        \midrule
            Checklist before export
            & \textbf{3.87}
            & 1.21
            & 3.71	
            & 1.40 
            & 0.16 \\
        \midrule
            Editing the agent responses
            & \textbf{4.29}
            & 0.76
            & 4.25	
            & 0.85 
            & 0.04 \\
        \bottomrule
    \end{tabular}
    \caption{Survey results (conducted after all three tasks given to both groups of annotators) about tool features rated by 7 internal annotators (IntAnno) and 24 external annotators (ExtAnno) based on their impact on the quality of created conversational data. A higher average value ($\mu$) indicates a bigger impact on conversational data quality (1 is ``No decrease [in data quality] at all'' and 5 is ``Extreme decrease in quality''.). $\mu$ Difference is the absolute difference between the two means from both groups of annotators. %\maeda{indicate the type of tool feature, i.e. whether it is for fixing the retrieval model output or the generator model output}
    }
    \label{tbl:survey}
\end{table}

%% file: discussion.tex
\section{Discussion}
\label{sec:discussion}

In this section, we distill insights from utilizing two groups of annotators with different feedback loops. %Our work is relevant to the CSCW community because 
We believe these practices can be applied to other tasks that utilize two populations for annotation, particularly when the task is complex and cannot be broken up like traditional micro-tasks~\cite{crowdsourcemicrotaskconv}. 
Our insights touch on various aspects of the overall conversational RAG data creation process, specifically task design (Section \ref{sec:insight-task} and \ref{sec:insight-task2}), tutorial materials (Section \ref{sec:insight-tutorial}), and tooling (Section \ref{sec:insight-tooling}). %\todo{explictly write out why our work and findings are of benefit to the CSCW community}

%\maeda{Expand on other dimensions: tooling and data, beyond the human aspects. The need to use a conversational interface, an online aspect (connected in real time to the model); you cannot have different groups of annotators doing different parts of the conversation in a disjoint manner (which we initially attempted). }

\subsection{Assign Creation and Review Tasks According to Annotator Group's Feedback Loop to Increase Throughput} \label{sec:insight-task}

Unlike micro-tasks, which are common in crowdsourced tasks~\cite{crowdsourcemicrotaskconv}, complex tasks like creating conversational RAG data cannot be broken up into smaller tasks without compromising the quality and the overall natural flow of the conversation~\cite{ragdatahard}. 
Moreover, the real-world limitation of an indirect type of feedback loop where the taskers cannot directly communicate with the crowd workers may decrease the quality of the created conversation. 
Given these two real-world constraints on task complexity and feedback annotator loop, we instead designed an additional, separate task, a review task, for the group of internal annotators who have a direct feedback loop with us (taskers). This resulted in the two-phase process as described in the data collection (Section \ref{sec:datacollection}), which we also observed in the literature, where creative tasks such as story writing cannot be broken into smaller tasks~\cite{mechanicalnovel2017,soylent2015}. Instead, creative tasks are ordered in a sequential flow of creation and review of writing tasks~\cite{soylent2015}.

Moreover, our study and survey results indicate that the direct and frequent feedback provided to the internal annotators helped them produce higher-quality conversations that are more complex, i.e., annotators attached a diverse set of passages to the conversation turns. However, it takes a considerable amount of time for them to create conversations, and there are fewer internal annotators. It took the internal annotators an average of 60-75 minutes to create a conversation and 30-45 minutes to review a conversation compared to the external annotators who took an average of 30-45 minutes to create a conversation. It is important to take advantage of both annotator pools to increase throughput. The findings in our review phase highlight that conversations can be continually improved. We propose a two-phase process where external annotators create the conversations, which will increase throughput and improve question diversity, and follow up with a review phase by the internal annotators to improve the quality of the conversation. 
A separate review task~\cite{reviewvsdoing2014}, or an expert review~\cite{annotationqualityassurance}, is not uncommon; however, most quality assurance strategies in crowdsourced tasks often employ filtering strategies~\cite{bostock2010crowdsourcedgraphical, annotationqualityassurance} (using qualification tests to filter out annotators), which is not possible in our setting. 
In our case, our group of expert annotators are characterized by their feedback loop, i.e., direct communication with the taskers as opposed to indirect communication through an intermediary, which has not been previously studied. 

\subsection{Provide Targeted Tasks to Increase Data Diversity}\label{sec:insight-task2}

In the crowdsourcing literature, there is a strong emphasis on breaking up tasks to reduce human error \cite{CrowdForge, annotationqualityassurance} but these best practices cannot be easily applied to non-decomposable macrotasks~\cite{nondecomposabletasks2018} such as creating complex RAG conversations. 
For instance, in the case of complex RAG conversations, we observed that the task of creating the initial seed question for the conversation can be performed as a separate task. %and the task of creating the remaining conversation. 
The benefit of breaking up the conversation creation task lies in the reduced burden on the annotators, especially when creating the initial seed question itself is an involved task, as it requires the annotator to come up with questions that are related to the corpora at hand and thus requires them to have also some level of insight into what kinds of questions may follow about such documents in the subsequent turns.

In our experience, task requesters can further constrain the design task of creating the initial seed question to be more \textit{targeted}. For instance, a targeted task for an initial seed question asks annotators to come up with questions that force the AI agent to respond with a \textit{clarifying question}, such as, \quotes{Which park are you referring to?}, when the initial seed question asks, \quotes{What time does the park close?}. 
Such targeted tasks reduce the chances of the annotator creating questions that are trivial and non-complex (that would be discarded since they do not lead to  long, meaningful and complex conversations). 
Moreover, targeted tasks can be given to crowd-like annotators, such as external annotators, while the task of creating the remaining turns can be given to the expert annotators (internal annotators). 
This design helps increase the diversity of the questions asked while maintaining high quality, which is also observed in prior work~\cite{snow-etal-2008-cheap} where the data utilized from crowdsourcing can perform as well as expert annotations as long as the task is broken down into less complex and more refined, smaller goals. 

%Prior work~\cite{snow-etal-2008-cheap} has shown that the lower quality data utilized from crowd-sourcing can perform as well as expert annotations in less complex tasks. We can utilize this approach in a more complex task by breaking down the task into smaller goals to increase the diversity of the questions asked while maintaining high quality. We have utilized the external annotators to create only the first question for the conversation and to create conversations with questions of a specific type that we are interested in exploring (e.g. under-specified or clarification questions) \todo{Double check if we defined the tasks for under-specified or clarification questions; we might need to drop this if we are not able to describe the targeted tasks (or define the tasks in Section 4.2.2) }. In these tasks, the external annotators can start the conversation and the internal annotators can then edit and continue the conversation to follow our metrics for complexity.
%\todo{Maeda to rewrite this section: utilizing the external annotators; which parts of the conversation creation task can be broken up: the creation of seed questions, which can be done by a crowd-type of annotators (in our case the external annotators)}

%\subsection{Design Tutorial Materials } 

\subsection{Utilize the Direct Feedback Loop to Improve the Annotation Guidelines} \label{sec:insight-tutorial}

In Section~\ref{sec:method}, we described the different kinds of instructions and feedback provided to both annotator groups. 
We observed that the combination of a slide-deck and a direct feedback loop, which includes access to instant messaging via Slack and weekly meeting sessions via Teams, with the internal annotators sufficed in training the internal annotators for creating complex and high-quality conversations. The direct feedback loop also gave us access and insight on areas of improvement for the tutorial materials. 
However, instructions and guideline materials given to the external annotators needed to be more comprehensive; in our case, we created video-based tutorial materials based on the slide-deck. 
Given the natural constraint of the indirect feedback loop, we also never received feedback regarding the quality of the tutorial materials from the external annotators. 
In the earlier phases of the data collection process, we observed that many of the data quality issues that came out of the final conversations created by the external annotators were the same issues the internal annotators had communicated with us directly. Examples of data quality issues included those that stemmed from grasping with how one would use the tool to do certain sub-tasks core to the overall conversation creation task, such as ensuring that the passages attached to the turns were diverse across the entirety of the conversation. 
Using feedback from the internal annotators, we refined the video tutorials and the annotation guidelines for the external annotators. This quick iterative process is important for spotting and resolving confusion to avoid both groups of annotators spending time creating conversations that would need to be discarded due to the confusion. 
Similar to how we utilized the internal annotators for a dedicated expert review task for quality assurance of the created conversations, we extended that direct feedback loop for improving the tutorial materials and instruction guidelines, which is especially important for the annotators with the indirect feedback loop. 
Previous work~\cite{Manam_Quinn_2018, Manam_Thomas_Quinn_2022} on using feedback to improve the task instructions often assume a direct feedback communication loop with all the annotators. In our work, we observed that feedback from a subset of the annotators, i.e. the internal annotators with direct feedback loop, can potentially improve task instructions and guidelines for the other subset of annotators with whom task requesters may not have direct communication.   

\subsection{Design Tooling to Facilitate Feedback that Cannot be Provided Directly } \label{sec:insight-tooling}

Conversational interfaces have been explored previously, but in different contexts, such as helping annotators effectively understand and complete crowdsourcing tasks or used for annotating complex artifacts such as code bases~\cite{annotationsindocs2022}, computational notebooks~\cite{Callisto2020}, conversations\cite{gilmartin-campbell-2016-capturing}, or even logs containing conversations between an AI voice agent and a customer~\cite{watsonxorders}. Conversational interfaces for creating conversations, specifically complex conversations, have been explored previously~\cite{ragdatahard, fadnis2025ragapheneragannotationplatform}, but the extent of best practices of design is not yet fully understood. In this work, we find that the collaborative aspect of the tooling (such as commenting on certain turns or attached passages of the created conversation) helped annotators understand which specific parts of the created conversation needed to be fixed. In short, tooling should be designed to facilitate some of the communication feedback that task requesters are unable to provide directly. 
%(1) the online aspect of the tooling facilitated a core part of creating conversations that flow naturally where subsequent turns made by the user depended on the context from the previous user/agent turns and (2) 
In our case, our survey results show that the annotators perceived that the hints that appeared in the tool as they created the conversation were key in ensuring the quality of the final conversation (see Section \ref{sec:annotator-perception-tool-feature}). 
Hints act as a form of feedback that task requesters would otherwise provide in direct feedback loops.  
In another example, the tooling feature in review mode where annotators could comment on specific parts of the created conversation, allowed other annotators to iterate and fix the mistakes or improve the quality of the created conversation. 
It was important for comments to be given at a fine-grain level, i.e., at a turn level, so that fixes can be made at that particular turn. 
Enabling expert annotators to leave fine-grained comments on created conversations helped improve the quality of the final conversation and also served as detailed feedback that was otherwise inaccessible for the external group of annotators that had an indirect feedback loop with the original task requesters.  
%\todo{Maeda to add: Also add how hints helped the annotators, which we found in our survey results in Section \ref{sec:annotator-perception-tool-feature}}

\section{Limitations and Biases}
\label{sec:limits}

We acknowledge that our study is not a controlled lab study; we report a real-world study, which primarily centers around our observations from tasking two groups of annotators on a complex, multi-layered task that has not been found in previous literature. Traditional quantitative methods popular in HCI literature also do not apply in our population due to the smaller pool of internal annotators compared to the number of external annotators. We also acknowledge that the task requesters (the authors of the paper) learned about the challenges of the task as we tasked the annotators. 

We also acknowledge that there are certain biases that may occur in each population due to their different incentives and tasks based on their feedback loop. The external annotators are paid per conversation, which encourages quantity but can impact quality, while the internal annotators are paid hourly, which discourages quantity and can improve quality.  

The review task is performed primarily by the internal annotators, and they may prefer their own conversations. We performed an analysis of the comments from the review of the external annotator conversations and found that 74\% of the conversations rejected by the internal annotators had at least one automated comment, indicating the external annotator did not follow the instructions for complexity. On the other hand, 60\% of the accepted conversations also had an automated comment, which indicates that the internal annotators were able to edit many of the conversations so they could be accepted. Some examples of comments from the annotators were: ``Rejected because there are no edits at all." [Int5], ``Three of the four questions were unanswerable, so I rejected the conversation." [Int6], ``I rejected the conversation because of the lack of diversity and flow. 3 of the question's answers were not found in the selected paragraphs. " [Int6], ``I reject this conversation since the first three questions are all asking about the same difference" [Int2]. These comments highlight either a lack of complexity (e.g., low passage diversity or no edits) and/or conversation flow (e.g., too many similar questions), which are important qualities that each conversation must have.

%% file: conclusion.tex
\section{Conclusion}
\label{sec:conclusion}

In conclusion, our longitudinal study explores the impact of different feedback loops for creating multi-turn RAG conversations across a three-phase task during the \studyduration of: pilot, create and review conversations and an accompanying survey. We find that a closer feedback loop with a small population of internal annotators, such as hands on meetings and Slack channels helped quickly refine guidelines for both populations and resulted in annotators that were experts in creating higher quality conversations according to our metrics. The larger population of external annotators with a larger feedback loop presents advantages in more diverse output and higher quantity of data. Our key takeaways for managing two population groups for complex annotation tasks are 1) utilize internal group to quickly refine guidelines that can be comprehensive to aid both groups, 2) give targeted tasks that are less complex and can be successfully completed by the external annotators, and 3) give the tasks completed by the external group to the internal annotators for review, editing and completion of tasks.